\documentclass[fleqn,usenatbib]{mnras}

\usepackage{newtxtext,newtxmath}
\usepackage[T1]{fontenc}

\DeclareRobustCommand{\VAN}[3]{#2}
\let\VANthebibliography\thebibliography
\def\thebibliography{\DeclareRobustCommand{\VAN}[3]{##3}\VANthebibliography}

\usepackage{graphicx}	% Including figure files
\usepackage{amsmath}	% Advanced maths commands
\usepackage{multirow}
\usepackage{ragged2e}
\usepackage{pdflscape}
\usepackage[flushleft]{threeparttable}
\usepackage{caption}

\newcommand{\cii}{[C\,{\sc ii}]}

\newcommand{\oi}{[O\,{\sc i}]}
\newcommand{\nii}{[N\,{\sc ii}]}

\newcommand{\oii}{[O\,{\sc ii}]}
\newcommand{\oiii}{[O\,{\sc iii}]}
\newcommand{\neiii}{[Ne\,{\sc iii}]}

\newcommand{\oiiil}{[O\,{\sc iii}] 88\,$\mu{\rm m}$}
\newcommand{\ciil}{[C\,{\sc ii}] 158\,$\mu{\rm m}$}

\title[Dust Build-up at $z\sim7$]{REBELS-IFU: Dust Build-up in Massive Galaxies at Redshift 7}

\author[H.S.B. Algera et al.]{Hiddo S. B. Algera$^{1,2,3}$\thanks{E-mail: hsbalgera@asiaa.sinica.edu.tw}, 
Lucie Rowland$^{4}$,
Mauro Stefanon$^{5,6}$, 
Marco Palla$^{7,8}$, 
Laura Sommovigo$^{9}$, 
\newauthor
Hanae Inami$^{2}$, 
Rychard Bouwens$^{4}$,
Manuel Aravena$^{10,11}$, 
Rebecca A. A. Bowler$^{12}$, 
Pratika Dayal$^{13}$, 
\newauthor
Ilse De Looze$^{14}$,
Andrea Ferrara$^{15}$, 
Rebecca Fisher$^{12}$, 
Luca Graziani$^{16,17,18}$, 
Cindy Gulis$^{19}$,
\newauthor
Kasper Heintz$^{20,21,22}$, 
Jacqueline Hodge$^{4}$, 
Andr\'{e}s Laza-Ramos$^{5}$,
Ivana van Leeuwen$^{4}$, 
Andrea Pallottini$^{15}$, 
\newauthor
Siân Phillips$^{23}$, 
Sander Schouws$^{4}$, 
Renske Smit$^{23}$, 
Daniel P.\ Stark$^{24}$, 
and Paul van der Werf$^{4}$ \\
$^{1}$Institute of Astronomy and Astrophysics, Academia Sinica, 11F of Astronomy-Mathematics Building, No.1, Sec. 4, Roosevelt Rd, Taipei 106319, Taiwan, R.O.C. \\
$^{2}$Hiroshima Astrophysical Science Center, Hiroshima University, 1-3-1 Kagamiyama, Higashi-Hiroshima, Hiroshima 739-8526, Japan \\
$^{3}$National Astronomical Observatory of Japan, 2-21-1, Osawa, Mitaka, Tokyo, Japan \\
$^{4}$Leiden Observatory, Leiden University, P.O. Box 9513, 2300 RA Leiden, The Netherlands \\
$^{5}$Departament d’Astronomia i Astrofìsica, Universitat de València, C. Dr. Moliner 50, E-46100 Burjassot, València, Spain \\
$^{6}$Unidad Asociada CSIC ”Grupo de Astrofísica Extragaláctica y Cosmología” (Instituto de Física de Cantabria - Universitat de València), Spain \\
$^{7}$Dipartimento di Fisica e Astronomia “Augusto Righi”, Alma Mater Studiorum, Università di Bologna, Via Gobetti 93/2, 40129 Bologna, Italy \\
$^{8}$INAF – Osservatorio di Astrofisica e Scienza dello Spazio di Bologna, Via Gobetti 93/3, 40129 Bologna, Italy \\
$^{9}$Center for Computational Astrophysics, Flatiron Institute, 162 5th Avenue,
New York, NY 10010, USA \\
$^{10}$Instituto de Estudios Astrof\'{\i}sicos, Facultad de Ingenier\'{\i}a y Ciencias, Universidad Diego Portales, Av. Ej\'ercito 441, Santiago, Chile \\
$^{11}$Millenium Nucleus for Galaxies (MINGAL) \\
$^{12}$Jodrell Bank Centre for Astrophysics, Department of Physics and Astronomy, School of Natural Sciences, The University of Manchester, Manchester M13 9PL, UK \\
$^{13}$Kapteyn Astronomical Institute, University of Groningen, P.O. Box 800, 9700 AV Groningen, The Netherlands \\
$^{14}$Sterrenkundig Observatorium, Ghent University, Krijgslaan 281 - S9, 9000 Gent, Belgium \\
$^{15}$Scuola Normale Superiore, Piazza dei Cavalieri 7, I-56126 Pisa, Italy \\
$^{16}$Dipartimento di Fisica, Sapienza, Università di Roma, Piazzale Aldo Moro 5, 00185 Roma, Italy \\
$^{17}$INFN, Sezione di Roma I, Piazzale Aldo Moro 2, 00185 Roma, Italy \\
$^{18}$INAF/Osservatorio Astronomico di Roma, Via di Frascati 33, 00078 Monte Porzio Catone, Italy \\
$^{19}$Department of Astronomy, the Pennsylvania State University, 525 Davey Lab, University Park, PA 16802 \\
$^{20}$Cosmic Dawn Center (DAWN), Denmark \\
$^{21}$Niels Bohr Institute, University of Copenhagen, Jagtvej 128, DK-2200 Copenhagen N, Denmark \\
$^{22}$Department of Astronomy, University of Geneva, Chemin Pegasi 51, 1290 Versoix, Switzerland \\
$^{23}$Astrophysics Research Institute, Liverpool John Moores University, 146 Brownlow Hill, Liverpool L3 5RF, UK\\
$^{24}$Department of Astronomy, University of California, Berkeley, Berkeley, CA 94720, USA 
}

\date{Accepted XXX. Received YYY; in original form ZZZ}

\pubyear{2025}
\begin{document}
\label{firstpage}
\pagerange{\pageref{firstpage}--\pageref{lastpage}}
\maketitle

\begin{abstract}
In recent years, observations with the \textit{JWST} have started to map out the rapid metal enrichment of the early Universe, while (sub)millimeter observations have simultaneously begun to reveal the ubiquity of dust beyond $z\gtrsim6$. However, the pathways that led to the assembly of early dust reservoirs remain poorly quantified, and require pushing our understanding of key scaling relations between dust, gas and metals into the early Universe. We investigate the dust build-up in twelve $6.5 \lesssim z \lesssim 7.7$ galaxies drawn from the REBELS survey that benefit from (i) \textit{JWST}/NIRSpec strong-line metallicity measurements, (ii) ALMA \cii{}-based redshifts and gas masses, and (iii) dust masses from single- or multi-band ALMA continuum observations. Combining these measurements, we investigate the dust-to-gas (DtG), dust-to-metal (DtM), and dust-to-stellar mass (DtS) ratios of our sample as a function of metallicity. While our analysis is limited by systematic uncertainties related to the \cii{}-to-H$_2$ conversion factor and dust temperature, we explore a wide range of possible values, and carefully assess their impact on our results. Under a fiducial set of assumptions, we find an average $\log(\mathrm{DtG}) = -3.02 \pm 0.23$, only slightly below that of local metal-rich galaxies. On the other hand, at fixed metallicity our average $\log(\mathrm{DtS}) = -2.15 \pm 0.42$ is significantly larger than that of low-redshift galaxies. Finally, through a comparison to various theoretical models of high-redshift dust production, we find that assembling the dust reservoirs in massive galaxies at $z\approx7$ likely requires the combination of rapid supernova enrichment and efficient ISM dust growth.

\end{abstract}


%%%%%%%%%%%%%%%%%%%%%%%%%%%%%%%%%%%%%%%%%%%%%%%%%%

%%%%%%%%%%%%%%%%% BODY OF PAPER %%%%%%%%%%%%%%%%%%

\section{Introduction}
\label{sec:introduction}
In recent years, the Atacama Large Millimeter/submillimeter Array (ALMA) has revealed the widespread presence of massive dust reservoirs in galaxies in the Epoch of Reionization (EoR; e.g., \citealt{riechers2013,watson2015,bowler2018,bowler2022,bowler2024,marrone2018,zavala2018,hashimoto2019,tamura2019,tamura2023,harikane2020,bakx2021,bakx2024,fudamoto2021,sugahara2021,akins2022,inami2022,schouws2022,witstok2022,witstok2023,algera2023,algera2024,algera2024b,decarli2023,hygate2023,tripodi2023,tripodi2024,mitsuhashi2024,valentino2024,vanleeuwen2024}).\footnote{The exact timeline of reionization has not yet been fully mapped out, but the latest observational estimates place its endpoint at $z\approx5.5$ (e.g., \citealt{bosman2022,spina2024}). In this work, however, we simply refer to the EoR as $z > 6$, when reionization was mostly complete.} In particular, ALMA has unambiguously revealed the presence of dust as early as $z=8.31$, corresponding to just $0.6\,$Gyr after the Big Bang \citep{tamura2019,tamura2023,bakx2020}. At even earlier cosmic epochs, the \textit{James Webb Space Telescope} (JWST) has indirectly inferred the presence of dust through its effect on the observed photometry and spectra of distant galaxies, suggesting modest levels of dust attenuation in some UV-luminous galaxies even at $z\gtrsim9$ (e.g., \citealt{cullen2023,boyett2024,carniani2024,ferrara2024}). 

These observations suggest early galaxies undergo rapid chemical enrichment, followed by a similarly swift build-up of dust. Several studies have attempted to quantify the dust masses ($M_\mathrm{dust}$) of high-redshift galaxies through multi-band ALMA observations, resulting in dust masses as high as $M_\mathrm{dust} \sim 10^{7 -8}\,M_\odot$ by $z\approx7$ \citep{hashimoto2019,bakx2021,akins2022,witstok2022,witstok2023,algera2024}. For larger galaxy samples, dust masses have been estimated based on a single-band ALMA detection, which requires the assumption of an underlying dust spectral energy distribution -- in particular the dust temperature ($T_\mathrm{dust}$; e.g., \citealt{sommovigo2022,sommovigo2022b}). While more uncertain, these single-band ALMA studies infer comparable dust masses to works benefiting from multi-band ALMA photometry (e.g., \citealt{ferrara2022,schouws2022,sommovigo2022}). 

To quantify the dust enrichment of distant galaxies, inferred dust masses are generally compared to galaxy stellar masses, suggesting typical ratios of $M_\mathrm{dust} / M_\star \sim 0.001 - 0.01$ \citep{witstok2023,algera2024}. Supernovae (SNe) have long been considered the main dust producers at high redshift (e.g, \citealt{todini2001}), owing to the short timescales on which they can begin to enrich the interstellar medium (ISM). As such, this dust-to-stellar mass ratio is often converted into an effective SN dust yield $y_\mathrm{SN}$, under the assumption of some initial mass function (IMF; e.g., \citealt{michalowski2015,lesniewska2019}). Given a fiducial \citet{chabrier2003} IMF, a ratio of $M_\mathrm{dust}/M_\star \sim 10^{-2}$ corresponds to a dust yield of approximately $y_\mathrm{SN} \approx 0.8\,M_\odot\,\mathrm{SN}^{-1}$.

Observationally, SN yields have been estimated in the local Universe, suggesting a wide range of values (see e.g., \citealt{sarangi2018} and  \citealt{schneider2024} for reviews). However, these yields are plagued by the unknown level of dust destruction due to the SN reverse shock (e.g., \citealt{bianchi2007,gall2018,micelotta2018,kirchschlager2019}), which could range anywhere from a few to nearly a hundred per cent (e.g., \citealt{nozawa2007,martinez-gonzalez2019,slavin2020,kirchschlager2023}). In an attempt to circumvent this uncertainty, \citet{galliano2021} perform an extensive study of the combined dust, gas and metal contents of $z=0$ galaxies, and empirically determine the effective SN yield post-reverse-shock destruction, suggesting $y_\mathrm{SN} \lesssim 0.1\,M_\odot\,\mathrm{SN}^{-1}$. If this low yield is also typical for high-redshift galaxies, SNe alone are likely insufficient to account for all the observed dust in the early Universe (e.g., \citealt{michalowski2015,lesniewska2019,algera2024b}). However, utilizing a different local galaxy sample and modeling approach, \citet{delooze2020} infer significantly higher SN yields, highlighting that such empirical approaches remain limited by degeneracies between the various dust production and destruction processes (see also \citealt{calura2023,park2024}). 

If SNe are indeed not sufficient to produce all the observed early dust, additional mechanisms must be at play. In the low-redshift Universe, Asymptotic Giant Branch (AGB) stars are known to be important sources of dust (e.g., \citealt{valiante2009}). The most massive AGB stars have lifetimes as short as $\sim40\,$Myr (e.g., \citealt{dellagli2019}), which may be shorter than the average ages of massive galaxies at $z\approx7$ \citep{whitler2023}. Regardless, given their long typical timescales, the contribution of AGB stars to global dust production is generally assumed to be subdominant at $z\gtrsim6$ \citep{mancini2015,lesniewska2019,dayal2022}. Similarly, other more `exotic' methods of dust production, such as Wolf-Rayet stars or quasar winds, are too believed to be unimportant at these early times \citep{schneider2024}. 

Instead, dust growth through metal accretion in the ISM is a more likely production pathway competing with SNe as the dominant mechanism of dust build-up at high redshift. This process requires some prior dust and metal enrichment -- likely from the aforementioned SNe -- after which gas-phase metals can deplete onto existing dust grains thereby increasing the total dust mass (e.g., \citealt{hirashita2012}). ISM dust growth is known to be particularly efficient in dense media (e.g., \citealt{weingartner1999,asano2013,roman-duval2014,zhukovska2016}), and given that high-redshift galaxies appear significantly denser than local galaxies (e.g., \citealt{isobe2023,ono2023}), it could be an efficient dust production pathway in the early Universe. On the other hand, the strong radiation fields and turbulence at high redshift may also limit the effectiveness of dust growth in the ISM and enhance dust destruction processes (e.g., \citealt{hirashita2009,ferrara2016}). 

From a theoretical perspective, various (semi-)analytical models and hydrodynamical simulations of high-redshift galaxies are now equipped with a framework tracking the production and destruction of dust (e.g., \citealt{mckinnon2016,mckinnon2017,popping2017,li2019,vijayan2019,graziani2020,triani2020,dayal2022,esmerian2022,esmerian2024,parente2022,dicesare2023,lewis2023,lower2023,lower2024,mauerhofer2023,choban2024,garaldi2024,matsumoto2024,narayanan2024,nikopoulos2024,palla2024,sawant2024}). These models and simulations thus predict the dust enrichment of the Universe both across time, and as a function of galaxy parameters, such as metallicity and stellar mass. Nevertheless, some of the key parameters characterizing dust growth, such as the SN yield or ISM growth timescale, are essentially unconstrained and often highly degenerate parameters in these dust production frameworks, and various combinations that match the limited number of observations available at early cosmic epochs can be devised (e.g., \citealt{esmerian2024,palla2024}). As such, testing their various predictions requires observations that target new regions of the parameter space, for example by extending observations to galaxies of lower dust and stellar masses (e.g., \citealt{valentino2024}), and by tying in information of galaxy metallicities (e.g., \citealt{heintz2023}).

With the advent of the \textit{JWST} it has now become possible to measure the gas-phase metallicities of large samples of distant galaxies (e.g., \citealt{heintz2023b,nakajima2023,chemerynska2024,curti2024,sanders2024,shapley2024}). At the same time, the number of galaxies in the Epoch of Reionization with ALMA-based dust mass measurements continues to increase (e.g., \citealt{bakx2021,witstok2022,algera2024}). Comparing the dust contents of galaxies at various metallicities is crucial to understand the overall production pathways of dust, given that different mechanisms have a different metallicity-dependence. For example, the dust-to-gas ratio depends roughly linearly on metallicity for SN dust production, but increases more rapidly in metal-rich galaxies when dust growth in the ISM is included (e.g., \citealt{asano2013,remyruyer2014,zhukovska2014,galliano2021}). 

By constraining scaling relations between dust, gas and metals at high redshift, we may therefore begin to disentangle the various dust production avenues in the early Universe. In the pre-\textit{JWST} era, such studies have been carried out up to Cosmic Noon ($z\sim1-3$; e.g., \citealt{shapley2020,popping2023}). While limited to small samples and high metallicities ($\gtrsim0.5\,Z_\odot$), these works suggest the $z\sim2$ scaling relation between dust-to-gas ratio and metallicity to be consistent with the local relation. Moreover, studies of Damped Lyman-$\alpha$ Absorbers (DLAs) are able to probe the dust-to-gas vs.\ metallicity relation out to $z\sim5$ (e.g., \citealt{decia2016,peroux2020}), and also find agreement with the local trends, at least at the high metallicity end. With the \textit{JWST}, direct studies of the dust, gas and metal contents of galaxies are now possible all the way into the Epoch of Reionization (e.g., \citealt{heintz2023,fujimoto2024a,valentino2024}), and can begin to constrain these fundamental scaling relations at even earlier times, where dust has had significantly less time to assemble. To date, however, the number of reionization-era galaxies benefiting from both \textit{JWST}/NIRSpec spectroscopy and ALMA observations of their dust and gas contents remains severely limited, which has thus far prevented any statistical insights into dust production pathways in the $z > 6$ Universe.

To address this limitation, we investigate the joint dust and metal contents of 12 galaxies drawn from the ALMA Reionization Era Bright Emission Line Survey (REBELS; \citealt{bouwens2022}) that benefit from new \textit{JWST}/NIRSpec spectroscopy, in order to shed light on the dust build-up in massive galaxies in the EoR. In Section \ref{sec:observations}, we introduce the ALMA and \textit{JWST} observations of our sample, and in Section \ref{sec:methods} we discuss how their dust and metal contents are inferred. We present our results in Section \ref{sec:results}, and subsequently discuss them in the context of high-redshift dust production in Section \ref{sec:discussion}, before concluding in Section \ref{sec:conclusions}. Throughout this work, we assume a standard $\Lambda$CDM cosmology, with $H_0=70\,\text{km\,s}^{-1}\text{\,Mpc}^{-1}$, $\Omega_m=0.30$ and $\Omega_\Lambda=0.70$. We further adopt a \citet{chabrier2003} IMF.

\begin{table*}
    \def\arraystretch{1.25}
    \addtolength{\tabcolsep}{-3.5pt} % slightly reduce column spacing to make the table fit
    \centering
    \caption{Physical properties of the 12 REBELS galaxies with combined ALMA and \textit{JWST}/NIRSpec observations, a.k.a., the REBELS-IFU sample.}
    \label{tab:data}
    \begin{tabular}{cccccccccc}
        \hline\hline 
         REBELS ID & \textbf{$z_\text{\cii{}}$} & $\log(L_\text{\cii{}}/L_\odot)$ & $12 + \log(\mathrm{O/H})$ & 
         $\log(\mathrm{M}_\star / M_\odot)$ & 
         $\log(\mathrm{M}_\mathrm{dust} / M_\odot)$ & 
         $\log(\mathrm{M}_{\mathrm{H}_2} / M_\odot)$ & 
         $\log(\mathrm{M}_\mathrm{dust} / \mathrm{M}_{\mathrm{H}_2})$ & 
         $\log(\mathrm{M}_\mathrm{dust} / \mathrm{M}_Z)$ & 
         $\log(\mathrm{M}_\mathrm{dust} / \mathrm{M}_\star)$ \\
         (1) & (2) & (3) & (4) & (5) & (6) & (7) & (8) & (9) & (10) \\
        \hline 
REBELS-05 & $6.4963$ & $8.84_{-0.06}^{+0.05}$ & $8.51\pm0.16$ & $9.42_{-0.10}^{+0.10}$ & $7.09_{-0.28}^{+0.55}$ & $10.32_{-0.30}^{+0.31}$ & $-3.16_{-0.47}^{+0.56}$ & $-1.13_{-0.50}^{+0.58}$ & $-2.30_{-0.32}^{+0.53}$ \\
REBELS-08 & $6.7495$ & $8.87_{-0.07}^{+0.06}$ & $8.22\pm0.20$ & $9.33_{-0.06}^{+0.07}$ & $7.27_{-0.29}^{+0.58}$ & $10.35_{-0.30}^{+0.31}$ & $-3.01_{-0.47}^{+0.59}$ & $-0.68_{-0.52}^{+0.61}$ & $-2.04_{-0.31}^{+0.57}$ \\
REBELS-12 & $7.3459$ & $9.01_{-0.22}^{+0.15}$ & $8.23\pm0.13$ & $9.54_{-0.04}^{+0.03}$ & $7.27_{-0.29}^{+0.62}$ & $10.47_{-0.37}^{+0.36}$ & $-3.10_{-0.54}^{+0.67}$ & $-0.79_{-0.55}^{+0.68}$ & $-2.26_{-0.31}^{+0.61}$ \\
REBELS-14 & $7.0842$ & $8.57_{-0.15}^{+0.11}$ & $7.90\pm0.12$ & $9.24_{-0.06}^{+0.07}$ & $7.07_{-0.29}^{+0.60}$ & $10.03_{-0.33}^{+0.33}$ & $-2.89_{-0.50}^{+0.62}$ & $-0.24_{-0.52}^{+0.62}$ & $-2.15_{-0.32}^{+0.59}$ \\
REBELS-15 & $6.8752$ & $8.28_{-0.11}^{+0.09}$ & $7.78\pm0.30$ & $9.31_{-0.01}^{+0.02}$ & $<7.10$ & $9.75_{-0.31}^{+0.31}$ & $< -2.58$ & $< 0.19$ & $< -2.21$ \\
REBELS-18 & $7.6750$ & $9.03_{-0.04}^{+0.03}$ & $8.50\pm0.13$ & $9.71_{-0.04}^{+0.06}$ & $7.19_{-0.29}^{+0.64}$ & $10.52_{-0.30}^{+0.31}$ & $-3.26_{-0.49}^{+0.63}$ & $-1.21_{-0.51}^{+0.64}$ & $-2.52_{-0.31}^{+0.62}$ \\
REBELS-25$^\dagger$ & $7.3065$ & $9.20_{-0.03}^{+0.03}$ & $8.62\pm0.17$ & $9.30_{-0.14}^{+0.12}$ & $8.20_{-0.40}^{+0.60}$ & $10.69_{-0.30}^{+0.31}$ & $-2.45_{-0.54}^{+0.63}$ & $-0.53_{-0.57}^{+0.65}$ & $-1.07_{-0.44}^{+0.60}$ \\
REBELS-29 & $6.6847$ & $8.74_{-0.08}^{+0.07}$ & $8.73\pm0.15$ & $9.69_{-0.05}^{+0.07}$ & $7.05_{-0.28}^{+0.56}$ & $10.22_{-0.30}^{+0.31}$ & $-3.11_{-0.47}^{+0.57}$ & $-1.29_{-0.50}^{+0.59}$ & $-2.63_{-0.30}^{+0.55}$ \\
REBELS-32 & $6.7290$ & $8.89_{-0.05}^{+0.04}$ & $8.48\pm0.13$ & $9.57_{-0.08}^{+0.10}$ & $7.05_{-0.28}^{+0.57}$ & $10.38_{-0.30}^{+0.31}$ & $-3.26_{-0.46}^{+0.57}$ & $-1.19_{-0.49}^{+0.58}$ & $-2.50_{-0.32}^{+0.55}$ \\
REBELS-34 & $6.6335$ & $8.84_{-0.15}^{+0.11}$ & $8.33\pm0.29$ & $9.45_{-0.02}^{+0.03}$ & $<7.14$ & $10.31_{-0.33}^{+0.33}$ & $< -3.10$ & $< -0.88$ & $< -2.30$ \\
REBELS-38$^\dagger$ & $6.5770$ & $9.23_{-0.04}^{+0.04}$ & $8.28\pm0.18$ & $9.75_{-0.06}^{+0.09}$ & $7.90_{-0.35}^{+0.39}$ & $10.72_{-0.30}^{+0.31}$ & $-2.81_{-0.48}^{+0.48}$ & $-0.55_{-0.51}^{+0.52}$ & $-1.87_{-0.36}^{+0.39}$ \\
REBELS-39 & $6.8449$ & $8.90_{-0.17}^{+0.12}$ & $8.02\pm0.29$ & $9.35_{-0.08}^{+0.09}$ & $7.17_{-0.29}^{+0.58}$ & $10.37_{-0.34}^{+0.34}$ & $-3.12_{-0.50}^{+0.61}$ & $-0.59_{-0.59}^{+0.67}$ & $-2.16_{-0.32}^{+0.56}$ \\
\hline 
\end{tabular}
    \raggedright \justify
    \textbf{Notes:} Col.\ (1): Galaxy identifier. Col.\ (2): \ciil{}-based spectroscopic redshifts from S.\ Schouws et al.\ (in prep). Col.\ (3): \cii{} luminosities from S.\ Schouws et al.\ (in prep). Col.\ (4): Metallicities, expressed as oxygen abundances, from \citet{rowland2025}. Col.\ (5) Stellar masses from M.\ Stefanon et al.\ (in prep). Col.\ (6): Dust masses. Col.\ (7): \cii{}-based molecular gas masses, assuming the \citet{zanella2018} conversion. Col.\ (8): dust-to-gas ratio. Col.\ (9): dust-to-metal ratio. Col.\ (10): dust-to-stellar mass ratio. $\dagger$ Dust masses obtained from multi-band ALMA photometry \citep{algera2024,algera2024b}.
\end{table*}

\section{Data}\label{sec:observations}

\subsection{ALMA data}
Our sample consists of 12 galaxies drawn from the Cycle 7 ALMA Large Program REBELS, described in detail in \citet{bouwens2022}. REBELS -- following the approach of its successful pilot studies \citep{smit2018,schouws2022,schouws2023} -- carried out ALMA spectral scans targeting the \ciil{} line and underlying dust continuum emission in 36 UV-luminous galaxies with photometric redshifts $z_\mathrm{phot} > 6.5$.\footnote{An additional four galaxies were targeted in \oiiil{} and underlying continuum emission, although none were detected in \oiii{}; see \citet{bouwens2022} and \citet{vanleeuwen2025}.} For details on the calibration and imaging, we refer the reader to \citet{bouwens2022} and \citet{inami2022}. In total, 25 galaxies were detected in \cii{} emission (Schouws et al.\ in preparation), of which 15 were also detected in dust continuum \citep{inami2022}. The \cii{} fluxes were measured through a 2D Gaussian fit to the moment-0 maps, collapsed across $2\times$ the line full-width at half maximum (Schouws et al.\ in preparation; see also \citealt{schouws2023}). The dust continuum flux densities are the peak intensities obtained from {\sc{CASA}} task {\sc{imfit}}, as detailed in \citet{inami2022}.\footnote{Only for REBELS-25 the integrated flux density was adopted in \citet{inami2022}, though we here utilize the peak flux density measurements extracted in tapered continuum images for this galaxy by \citet{algera2024b}; see also Section \ref{sec:methodsDustMass}.}

In this work, we focus on the 12 \cii{}-detected REBELS galaxies with follow-up \textit{JWST}/NIRSpec Integral Field Unit (IFU) spectroscopy, hereafter referred to as the `REBELS-IFU' sample (see below). These galaxies have spectroscopic redshifts in the range $6.5 \lesssim z \lesssim 7.7$, \cii{} luminosities between $\log(L_\text{\cii} / L_\odot) = 8.2 - 9.3$, and stellar masses spanning $\log(M_\star/M_\odot) = 9.2 - 9.8$ (Table \ref{tab:data}). Moreover, 10/12 galaxies are detected in ALMA Band 6 continuum emission, probing rest-frame $160\,\mu\mathrm{m}$ \citep{inami2022}. For the remaining two galaxies, we adopt $3\sigma$ upper limits on their continuum emission, where $\sigma$ is the root-mean-square (rms) noise in the naturally-weighted Band 6 continuum image.\footnote{One of these two galaxies is REBELS-34, for which \citet{bowler2022} obtain a $4.5\sigma$ continuum detection using more sensitive ALMA observations (REBELS-34 is known as ID 169850 in their work). However, we here focus on the continuum observations taken as part of REBELS, following e.g., \citet{inami2022} and \citet{bouwens2022}; our $3\sigma$ upper limit is fully consistent with the faint continuum detection in \citet{bowler2022}.}

\subsection{JWST data}
Twelve of the most \cii{}-luminous REBELS galaxies were followed up in a Cycle 1 \textit{JWST} program (GO 1626 [PI Stefanon] targeted 11/12 galaxies, while REBELS-18 was observed as part of GO 2659 [PI Weaver]). These observations made use of low-resolution ($R\sim100$) \textit{JWST}/NIRSpec IFU prism spectroscopy, covering a $3'' \times 3''$ area on the sky across a wavelength range of $0.6 - 5.3\,\mu\mathrm{m}$. This setup ensures coverage of many of the key rest-frame UV and optical emission lines, such as Lyman-$\alpha$, the [OII]$_{3727,29}$ doublet, H$\beta$ and the [OIII]$_{4959,5007}$ lines. For the 8/12 sources at $6.5 \lesssim z\lesssim7$, we moreover cover the H$\alpha$ line, while at higher redshift this line moves out of the NIRSpec wavelength range. Details on the NIRSpec data reduction, emission line extraction, and metallicity measurements are presented in papers accompanying this work (\citealt{rowland2025}; Stefanon et al.\ in preparation), while we provide a brief summary in Section \ref{sec:methodsMetallicity}.

While overall representative of the 25 \cii{}-detected galaxies within REBELS, we note that the REBELS-IFU targets were selected with a preference towards \cii{}-luminous and continuum-detected sources (Stefanon et al.\ in preparation). This may bias their selection towards the most metal- and dust-rich galaxies among the parent sample, and should be borne in mind when discussing their dust build-up in Section \ref{sec:discussion}.

\section{Methods}
\label{sec:methods}

\subsection{Dust mass measurements}
\label{sec:methodsDustMass}

Among the ten dust continuum-detected galaxies in the REBELS-IFU sample, two have been detected in more than one ALMA band: REBELS-25 has been targeted in six unique bands, and as such has a well-constrained dust mass, temperature and emissivity index ($\beta_\mathrm{IR}$; \citealt{algera2024b}). REBELS-38 has been detected in ALMA Bands 6 and 8, enabling a dual-band dust mass measurement \citep{algera2024}. While REBELS-12 was also followed up in Band 8, it was not detected, and the data were of insufficient depth to facilitate a robust dust mass measurement \citep{algera2024}. This thus leaves eight dust-detected galaxies for which only a single-band dust mass measurement is available. 

Dust masses for the continuum-detected REBELS galaxies have previously been determined by \citet{sommovigo2022,sommovigo2022b}, leveraging the \cii{}-based method introduced in \citet{sommovigo2021}. Briefly, this method adopts \cii{} as both a gas and a star formation rate (SFR) tracer, and assumes a priori that the dust-to-gas ratio depends linearly on metallicity. Combined with a single-band continuum detection, these assumptions enable an estimate of the dust mass. 

Because of the implicit assumption on the metallicity-dependence of the dust-to-gas ratio, however, we opt not to use the \citet{sommovigo2022} dust masses in this work, as this is one of the scaling relations that we are interested in constraining directly. Instead, we assume a fixed dust temperature of $T_\mathrm{dust} = 45 \pm 15\,$K for the 10 galaxies where we have only a single-band ALMA detection, or no continuum detection at all (REBELS-15, REBELS-34). In the dust mass determination, we furthermore adopt a fixed emissivity index ($\beta_\mathrm{IR}=2.0$) and opacity coefficient ($\kappa_0 = 10.41\,\mathrm{g}^{-1}\mathrm{cm}^2$ at $\nu_0=1900\,\mathrm{GHz}$), following e.g., \citet{ferrara2022}. \citet{algera2024,algera2024b} adopted the same dust model in their analysis of the two sources with multi-band ALMA continuum detections, REBELS-25 and REBELS-38, and we therefore adopt their measurements of the dust temperature and mass for these two galaxies directly. In the dust mass determinations, we furthermore account for heating by as well as attenuation against the CMB following \citet{dacunha2013}.

Focusing on the 8 single-band dust-detected galaxies in common between the REBELS-IFU and \citet{sommovigo2022} samples, we find good agreement between our dust masses with a mean offset of $\Delta\log(M_\mathrm{dust}) = 0.02 \pm 0.12\,\mathrm{dex}$, where the error represents the standard deviation. This is unsurprising, as the mean dust temperature of the 13 sources analyzed in \citet{sommovigo2022} is $T_\mathrm{dust}\approx46\,\mathrm{K}$, very similar to the value adopted in this work. We discuss the effect of our assumed dust temperature in further detail in Section \ref{sec:errorBudget}. 

\begin{figure*}
    \centering
    \includegraphics[width=1.0\textwidth]{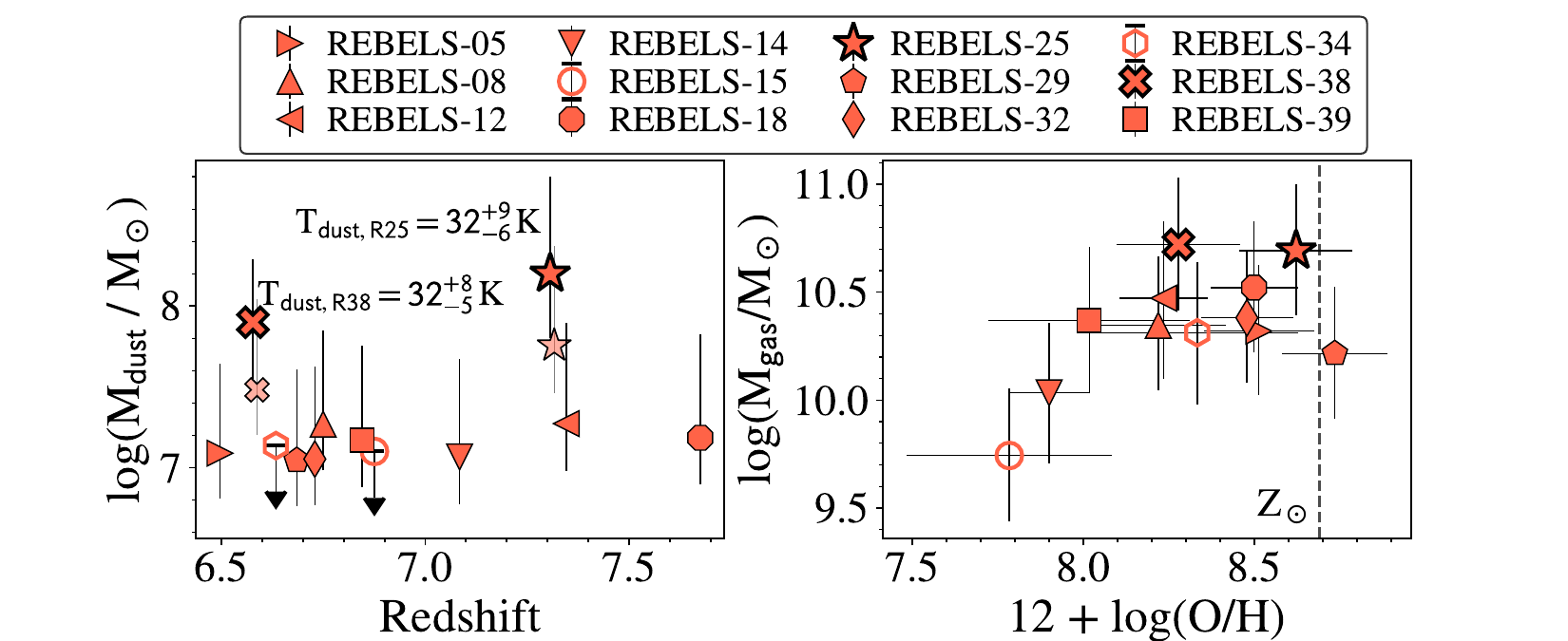}
    \caption{The physical properties of the 12 targets with combined \textit{JWST} + ALMA observations studied in this work. \textit{Left:} dust mass vs.\ redshift. We show the two galaxies for which we have multi-band dust mass estimates -- REBELS-25 and REBELS-38 \citep{algera2024,algera2024b} -- twice; once using the measured dust mass (large symbols with thick outline), and once assuming the same fiducial $T_\mathrm{dust} = 45 \pm 15\,\mathrm{K}$ as for the rest of the sample (fainter symbols; offset in redshift by $\Delta z = 0.01$ for clarity). The two REBELS-IFU targets without a continuum detection are shown as empty markers. For these, only upper limits on $M_\mathrm{dust}$ are available. \textit{Right:} \cii{}-based molecular gas mass vs.\ metallicity. The gas mass is inferred using the calibration from \citet{zanella2018}, who determined a mass-to-light ratio of $\alpha_\text{\cii{}} = 31 M_\odot\,L_\odot^{-1}$. The solar metallicity [$12 + \log(\mathrm{O/H})_\odot = 8.69$] is indicated via the vertical dashed line. Our $6.5 \lesssim z \lesssim 7.7$ sample spans $\sim1\,\mathrm{dex}$ in dust mass, gas mass and metallicity.}
    \label{fig:data}
\end{figure*}

We show the range of dust masses spanned by our sample, $\log(M_\mathrm{dust}/M_\odot) \approx 7.0 - 8.2$, as a function of redshift in the left panel of Fig.\ \ref{fig:data}. We note that, except for the two sources with multi-band ALMA detections, the dust masses of the continuum-detected sample span a relatively narrow range of $7.0 \lesssim \log(M_\mathrm{dust}/M_\odot) \lesssim 7.3$. This is partially because the two sources with multi-band ALMA follow-up are the most luminous among the REBELS sample at rest-frame $160\,\mu\mathrm{m}$, suggesting a high dust content, but mostly because we infer a lower dust temperature for them compared to what we assume for the rest of the sample (see Section \ref{sec:errorBudget}). Had we assumed the same $T_\mathrm{dust} = 45\pm15\,\mathrm{K}$ for REBELS-25 and REBELS-38, their estimated dust masses would be lower by $\sim0.4\,\mathrm{dex}$ (semi-transparent symbols in Fig.\ \ref{fig:data}), although still consistent with the multi-band measurements within the uncertainties.

\subsection{Gas mass measurements}
\label{sec:gasMassMeasurements}

To investigate the pathways of early dust build-up, the dust-to-gas (DtG) and dust-to-metal (DtM) ratios are particularly important quantities. We note that these are related via $\mathrm{DtM} = M_\mathrm{dust} / M_\mathrm{Z} = (M_\mathrm{dust} / M_\mathrm{gas}) \times (M_\mathrm{gas} / M_\mathrm{Z}) = \mathrm{DtG}/ Z$, where $M_Z = M_\mathrm{gas} \times Z$ is the total mass of gas-phase metals. Crucially, these quantities require knowledge of the total gas mass, $M_\mathrm{gas}$. This, in turn, is a combination of molecular and atomic gas ($M_\mathrm{gas} = M_{\mathrm{H}_2} + M_{\mathrm{HI}}$). Atomic gas is notoriously difficult to measure in the early Universe, and we therefore neglect it in our analysis.\footnote{We note, however, that the \textit{JWST} has enabled studies of Lyman-$\alpha$ damping wing absorption beyond $z\gtrsim7$ which provides information on atomic hydrogen along the line-of-sight (see e.g., \citealt{curtis-lake2023,heintz2024,umeda2024}).} We justify this assumption in Section \ref{sec:errorBudget}.

To infer molecular gas masses at lower redshifts, CO or [CI] are typically used as tracers (e.g., \citealt{hagimoto2023,friascastillo2024}). However, these lines are too faint to be detected in all but the brightest galaxies and quasars in the Epoch of Reionization (e.g., \citealt{strandet2017,ono2022,decarli2022,hashimoto2023,kaasinen2024}). Moreover, even if sufficiently bright, generally only the higher rotational transitions of CO are available. These higher-J lines typically do not trace the full, cold molecular gas reservoir, but instead a denser, warmer component (e.g., \citealt{vallini2018}). 

To homogeneously estimate the gas masses of our sample, we therefore adopt the \cii{} line as a tracer. In particular, we follow the work by \citet{aravena2024}, who previously investigated the molecular gas masses of REBELS galaxies, by adopting the conversion proposed by \citet{zanella2018}. \citeauthor{zanella2018} obtained a relation between \cii{} luminosity and molecular gas mass across a heterogeneous sample of local and intermediate-redshift ($z\sim0.5 - 6$) main-sequence galaxies and starbursts. We reproduce their relation here for reference:

\begin{align}
    \log \left(\frac{L_\text{\cii}}{L_\odot}\right) = -(1.28 \pm 0.21) + (0.98 \pm 0.02) \log\left(\frac{M_{\mathrm{H}_2}}{M_\odot}\right) \ .
    \label{eq:zanella}
\end{align}

\noindent Since this \cii{}-to-H$_2$ relation is nearly linear, we follow \citet{zanella2018} by expressing it as a mass-to-light ratio, evaluating the expression at the median luminosity of the REBELS-IFU sample, $\langle L_\text{\cii{}} \rangle = 10^{8.9}\,L_\odot$. This results in $\alpha_\text{\cii{}} = M_{\mathrm{H}_2} / \langle L_\text{\cii{}} \rangle \approx 31_{-15}^{+31}\,M_\odot\,L_\odot^{-1}$. We further note that \citet{zanella2018} do not find evidence that $\alpha_\text{\cii{}}$ varies with metallicity. While the subset of their sample for which metallicity measurements are available is relatively enriched ($Z \gtrsim 0.2\,Z_\odot$), this matches the metallicities inferred for the REBELS- IFU sample (Section \ref{sec:methodsMetallicity}).

We propagate the uncertainty on the \cii{}-to-H$_2$ conversion throughout our analysis, resulting in typical uncertainties on $M_{\mathrm{H}_2}$ of $\sim0.3\,\mathrm{dex}$ (Table \ref{tab:data}). We note that -- despite being adopted in numerous high-redshift studies (e.g., \citealt{dessauges-zavadsky2020,bethermin2023,aravena2024}) -- the \cii{}-to-H$_2$ conversion factor is poorly constrained, and we discuss this further in Section \ref{sec:errorBudget} (see also \citealt{casavecchia2024} for a recent discussion). While kinematically-inferred gas masses could be used to circumvent this uncertainty, the resolution of the REBELS \cii{} observations ($1.2 - 1.6''$) is insufficient for robust constraints on the dynamical states and sizes of our targets (see also the discussion in \citealt{aravena2024}). Consequently, these dynamical estimates are similarly (if not more) uncertain, and we do not use them in this work. However, Algera \& Herrera-Camus et al.\ (in preparation) find that, for two $z>5$ galaxies for which detailed kinematic modeling of their high-resolution \cii{} observations is possible, the \cii{}-based and dynamical gas mass estimates are in good agreement.

We furthermore note that our \cii{} luminosities (and thus molecular gas masses) are not corrected for attenuation against the CMB. While the exact magnitude of this effect depends on the gas temperature and density \citep[e.g.,][]{dacunha2013,lagache2018,kohandel2019}, it is likely to be small for the REBELS-IFU sample. On the basis of \nii{}$_{205}$ non-detections in three REBELS-IFU galaxies, \citet{fudamoto2025} argue that most of their \cii{} emission originates from photo-dissociation regions (PDRs), while \oi{}$_{145}$ detections in the same sample suggest high PDR densities of $\log(n/\mathrm{cm}^{-3}) > 3.5$. If these densities are also typical for the rest of our sample, no significant CMB attenuation is expected ($<10\%$) even if gas temperatures are modest \citep[e.g.,][]{lagache2018,kohandel2019}.

We show the inferred molecular gas masses of our sample in Fig.\ \ref{fig:data} (right panel), as a function of their rest-frame optical metallicity (Section \ref{sec:methodsMetallicity}). Our targets span roughly $1\,\mathrm{dex}$ in molecular gas mass, ranging from $\log(M_{\mathrm{H}_2} / M_\odot) = 9.7 - 10.7$. However, most sources are clustered around $\log(M_{\mathrm{H}_2} / M_\odot) \approx 10.4$, which is due to the parent REBELS sample essentially being \cii{}-selected (see also the discussion in \citealt{aravena2024}).

\subsection{Stellar mass measurements}
\label{sec:stellarMass}
Two separate studies derive stellar masses for the REBELS-IFU sample: Stefanon et al.\ (in preparation) and \citet{fisher2025}. We here adopt the former as fiducial, for consistency with the work on the mass-metallicity relation of REBELS-IFU galaxies by \citet{rowland2025}. While both Stefanon et al.\ and \citet{fisher2025} adopt a similar approach -- summarized below -- the key difference is that the latter adopt a flexible dust law in their analysis, while the masses used here adopt a fixed \citet{calzetti2000} dust attenuation curve. Regardless, \citet{fisher2025} find that most of the REBELS-IFU sources are consistent with the Calzetti curve, and the two mass measurements are therefore in good overall agreement.

To estimate the stellar masses, Stefanon et al.\ (in preparation) run \textsc{Bagpipes} \citep{carnall2018,carnall2019} on the extracted IFU spectra using the 2016 version of the \citet{bruzual2003} stellar population models.\footnote{Stefanon et al.\ (in preparation) adopt a \citet{kroupa2001} IMF, while we follow \citet{michalowski2015} by adopting a \citet{chabrier2003} IMF in our analysis of dust-to-stellar mass ratios in Section \ref{sec:discussionComparisonObservations}. However, stellar masses inferred using these two IMFs agree to within $\lesssim0.04\,\mathrm{dex}$ (e.g., \citealt{madau2014}), which is well within the observational uncertainties; as such, we do not correct for this.} The redshifts of the sources are fixed to the \cii{}-based values, and for the metallicities a Gaussian prior is adopted using the measurement and corresponding uncertainty from \citet{rowland2025} as the mean and standard deviation, respectively. Stefanon et al.\ provide fits with two distinct star formation histories (SFHs): either a constant SFH plus a single burst, or a non-parametric SFH. For consistency with \citet{rowland2025}, we adopt the non-parametric SFH as fiducial in our analysis. If we were to instead adopt the constant + burst SFH, the average stellar mass of our sample would slightly increase, by $\Delta \log(M_\star) = 0.10 \pm 0.13\,\mathrm{dex}$. However, this has no effect on our conclusions.

In modeling the non-parametric SFH, the formation time of the galaxy is set to $z=30$, and the SFH is divided into four distinct time bins. The width of the first two bins is fixed, and is defined to capture recent star formation in the galaxy across the last $0-3\,\mathrm{Myr}$ and $3-13\,\mathrm{Myr}$. The final two bins are equally divided in time back to the formation epoch. To link the different time bins, Stefanon et al.\ (in preparation) assume a continuity prior following, e.g., \citet{leja2019} and \citet{tacchella2022}. 

The stellar masses of the REBELS-IFU sample are presented in Table \ref{tab:data}, and span a range of $\log(M_\star/M_\odot) = 9.2 - 9.8$. We note that in some instances the masses have decreased with respect to the \citet{topping2022} values that were previously adopted in several REBELS publications (e.g., \citealt{sommovigo2022b,algera2023,aravena2024}). These differences are mainly due to the improved S/N and spectral sampling provided by the NIRSpec observations with respect to previous ground-based and \textit{Spitzer}/IRAC data, as well as the improved spatial resolution reducing contamination from neighboring sources (see Stefanon et al.\ in preparation for details).

\subsection{Metallicity measurements}
\label{sec:methodsMetallicity}

Gas-phase metallicities have been measured for our 12 targets by \citet{rowland2025}, and we refer to their work for details while summarizing the procedure in the following.\footnote{We use metallicity, denoted by $Z$, interchangeably with Oxygen abundance [denoted $12 + \log(\mathrm{O/H})$]. To convert between the two, we adopt solar abundance ratios \citep{asplund2009}.} Briefly, the metallicities are obtained from the $R23 = (\mathrm{[O\,III]}_{4959,5007} + \mathrm{[O\,II]}_{3727,29}) / \mathrm{H}\beta$, $R3 = \mathrm{[O\,III]}_{5007} / \mathrm{H}\beta$, or $O32 = \mathrm{[O\,III]}_{5007} / \mathrm{[O\,II]}_{3727,29}$ line ratios using the strong-line metallicity calibrations from \citet{sanders2024}. However, \citet{rowland2025} show that consistent results are obtained when other calibrations are adopted, such as those from \citet{nakajima2022}.  

To measure the gas-phase metallicities of the REBELS-IFU targets, we first extract the 1D spectra from the IFU cubes by summing all pixels showing significant emission (Fig.\ 1 in \citealt{rowland2025}). The stellar continuum is fitted with a third-order polynomial, and is subsequently subtracted. Following this, various emission lines are fitted, including but not limited to the $\mathrm{[O\,II]}_{3727,29}$ doublet --  unresolved, and therefore fitted with a single Gaussian -- H$\beta$, the $\mathrm{[O\,III]}_{4959,5007}$ doublet -- resolved and fitted with two Gaussians -- and H$\alpha$. The latter is deblended from the $\mathrm{[N\,II]}_{6548,84}$ doublet by a triple-Gaussian fit. 

For our targets at $z\lesssim7$ (8/12), we detect both the H$\beta$ and H$\alpha$ lines, and use the resulting Balmer decrement to correct the emission line fluxes for dust attenuation. At higher redshifts, the H$\alpha$ line moves out of the NIRSpec coverage, preventing robust dust corrections. While in theory the H$\gamma$/H$\beta$ ratio could be used to apply dust corrections, the H$\gamma$ line is blended with both the $\mathrm{[Fe\,II]}_{4360}$ and $\mathrm{[O\,III]}_{4363}$ lines, and is moreover only detected at low significance (see the discussion in \citealt{rowland2025}). As a result, we do not apply dust corrections to the emission line fluxes for our four targets at $z > 7$. Accordingly, we use a metallicity indicator insensitive to dust attenuation to infer their level of metal enrichment, as detailed below. \\

To determine the metallicities of the REBELS-IFU sample, we use the $R23$ calibration where possible, which is generally regarded as the most reliable strong-line indicator given its incorporation of two different ionization states of oxygen \citep{maiolino2019}. However, given the significant difference in wavelength between the $\mathrm{[O\,II]}_{3727,29}$ and $\mathrm{[O\,III]}_{4959,5007}$ doublets, the effects of dust extinction are potentially significant, and we therefore do not apply the $R23$ diagnostic to $z > 7$ sample. Moreover, the relation between $R23$ and metallicity is double-valued, and shows a maximum around an oxygen abundance of $12 + \log(\mathrm{O/H}) \approx 8$ \citep{sanders2024}. We therefore use the $O32$ and $Ne3O2$ = $\text{\neiii{}}_{3869}/\text{\oii{}}_{3727,29}$ diagnostics to decide whether to use the lower or upper branch of the $R23$-metallicity relation (and do the same for the double-valued $R3$ diagnostic, see below). While for the $z>7$ sample we cannot reliably dust-correct the $O32$ diagnostic, \citet{rowland2025} find that using the upper limits obtained from the dust-uncorrected line fluxes is already sufficient to break the degeneracy between the two branches. Moreover, consistent results are found when using the $Ne3O2$ diagnostic, which is not significantly affected by dust.

We note that three of our $z < 7$ sources -- REBELS-15, REBELS-34 and REBELS-39 -- have $R23$ (and $R3$) ratios in excess of the maximum value calibrated by \citet{sanders2024}, preventing a metallicity measurement using the $R23$ diagnostic. This could point towards the presence of an active galactic nucleus (AGN), although \citet{rowland2025} find that this cannot be conclusively established based on the inspection of several relevant emission line diagnostics. We therefore adopt the single-valued $O32$ calibration to infer the metallicities for these three sources. Finally, for the $z > 7$ sample, we adopt the $R3$ calibration, which makes use of two lines that are closely separated in wavelength and thus does not require a dust correction. 

Overall, our fiducial metallicity measurements thus use $R23$ ($O32$) for 5 (3) galaxies at $z < 7$, and $R3$ for $4$ galaxies at $z > 7$. However, in Appendix \ref{app:metallicity} we explore the effects of using only the $O32$ diagnostic, which is single-valued and available for the full REBELS-IFU sample, albeit also without dust corrections for the $z\gtrsim7$ sources. For these four galaxies, the $O32$ diagnostic therefore yields a lower limit on their metallicity. For the five galaxies at $z \lesssim 7$ for which we use the $R23$ calibration as fiducial, adopting $O32$ instead yields oxygen abundances that are, on average, $0.17 \pm 0.24\,\mathrm{dex}$ lower, which does not significantly impact our results. 

As a further consistency check, \citet{rowland2025} have also remeasured the metallicities of the REBELS-IFU sample using dust corrections obtained from SED-fitting, which yields the stellar V-band attenuation $A_{V,*}$. By scaling this to the nebular attenuation $A_{V,\mathrm{neb}}$ via $A_{V,*} = 0.44 A_{V,\mathrm{neb}}$ based on \citet{calzetti1994}, we recompute the metallicities for the four $z>7$ sources based on the dust-corrected $R23$ diagnostic. This yields metallicities that are, on average, $0.09\,\mathrm{dex}$ larger than our fiducial values. This is, however, fully consistent with the fiducial measurements within their uncertainties \citep[see also Section 3.1 in][]{rowland2025}.

We show the fiducial metallicities of our sample in the right panel of Fig.\ \ref{fig:data}. The REBELS-IFU targets span a range of $7.8 \lesssim 12 + \log(\mathrm{O/H}) \lesssim 8.7$ (corresponding to $\sim0.1 - 1.1\,Z_\odot$), though all but two of the 12 galaxies have $12 + \log(\mathrm{O/H}) > 8.0$. For reference, the solar metallicity and oxygen abundance are taken to be $Z_\odot = 0.014$ and $12 + \log(\mathrm{O/H}) = 8.69$, respectively, following \citet{asplund2009}. The uncertainty on the metallicities, obtained from propagating the errors on the relevant emission line fluxes, ranges from $0.1 - 0.3\,$dex (Table \ref{tab:data}). We note that this uncertainty does not include any systematic errors in the calibrations themselves (e.g., \citealt{kewley2008}). As such, the derived metallicity for any individual galaxy is likely to be more uncertain than suggested by the formal measurement errors. However, on average the REBELS-IFU galaxies are more metal-rich than the typical $z\sim7$ galaxy population, irrespective of which metallicity diagnostic is adopted (Section 4.2.1 in \citealt{rowland2025}).

\subsection{Error budget}
\label{sec:errorBudget}

\begin{figure}
    \centering
    \includegraphics[trim={0 1.0cm 0 0}, width=0.52\textwidth]{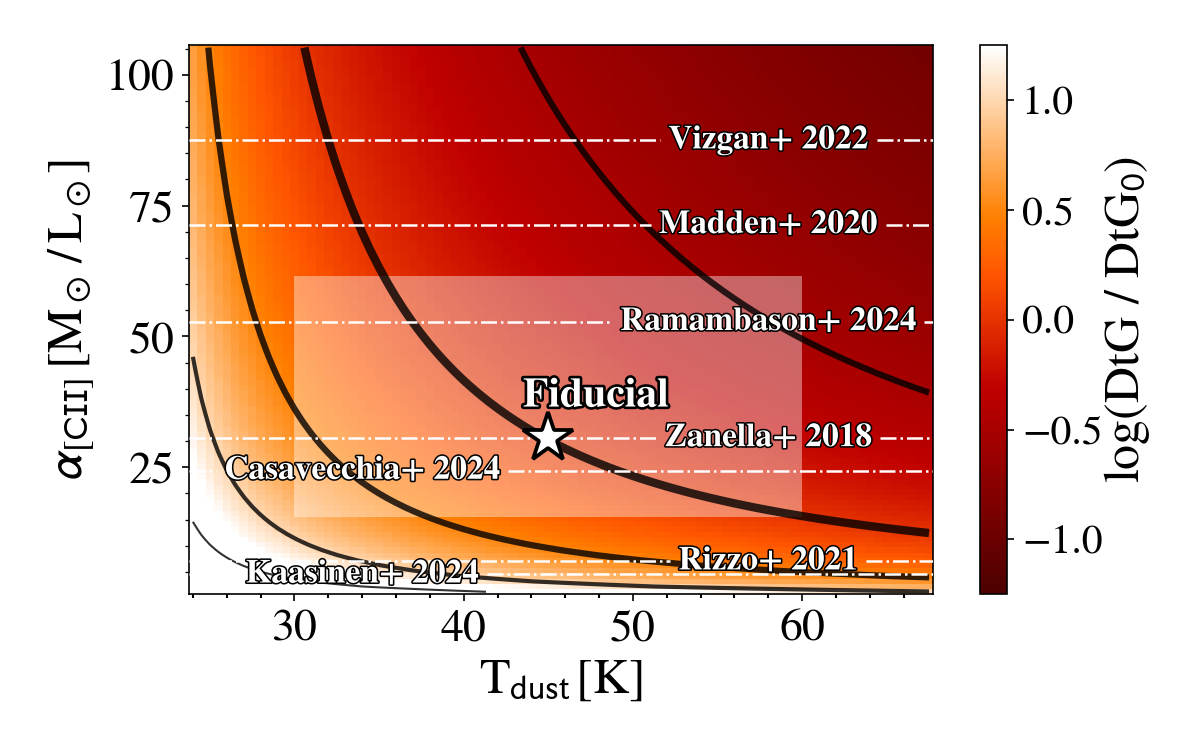}
    \caption{The effect of the adopted $\alpha_\text{\cii{}}$ and $T_\mathrm{dust}$ on the inferred dust-to-gas ratio of a hypothetical galaxy at the median redshift ($z=6.8$) and \cii{} luminosity [$\log(L_\text{\cii{}}/L_\odot) = 8.9$] of our sample. The thickest contour corresponds to $\log(\mathrm{DtG}/\mathrm{DtG}_0) = 0$, where the fiducial $\mathrm{DtG}_0$ is represented by the white star and assumes $T_\mathrm{dust} = 45\,$K and the \cii{}-to-H$_2$ conversion factor from \citet{zanella2018}. The white, shaded region shows the uncertainties adopted on the dust temperature ($\pm15\,\mathrm{K}$) and $\alpha_\text{\cii{}}$ (approximately $\pm0.3\,\mathrm{dex}$). Increasingly thinner contours show offsets of $\pm0.5, 1.0, 1.5\,\mathrm{dex}$ with respect to the fiducial DtG ratio. Various $\alpha_\text{\cii{}}$ determinations from the literature are furthermore overplotted \citep{madden2020,rizzo2021,vizgan2022,casavecchia2024,kaasinen2024,ramambason2024}.}
    \label{fig:grid}
\end{figure}

\subsubsection{Dust temperatures}

Prior to discussing the dust and gas properties of the REBELS-IFU sources in detail, it is worth re-iterating some of the key uncertainties in our analysis. First of all, the dust temperatures of 10/12 of our targets are unknown, and as detailed in Section \ref{sec:methodsDustMass}, we therefore assume a fixed temperature of $T_\mathrm{dust} = 45 \pm 15\,\mathrm{K}$. This value is similar to the average dust temperature obtained by \citet{witstok2023}, who determined $T_\mathrm{dust} = 53 \pm 13\,\mathrm{K}$ for a heterogeneous sample of $z\gtrsim4$ galaxies and quasars with multi-band ALMA photometry. Moreover, our adopted value is similar to the average temperature of $T_\mathrm{dust} = 41_{-9}^{+10}\,\mathrm{K}$ obtained by \citet{mitsuhashi2024} for a sample of UV-luminous galaxies at $z\sim6$ drawn from the SERENADE survey. The assumed dust temperature of $T_\mathrm{dust} = 45\,$K is, however, slightly higher than what was directly measured for REBELS-25 and REBELS-38 using multi-band dust continuum data ($T_\mathrm{dust} \approx 30 - 35\,$K; \citealt{algera2024,algera2024b}). 

We note that at lower redshifts, galaxy dust masses are commonly estimated using a fixed dust temperature of $T_\mathrm{dust} = 25\,\mathrm{K}$. This temperature is thought to be typical for their cold dust reservoir which accounts for most of the overall dust mass (e.g., \citealt{scoville2016,pozzi2021}). However, this is likely not appropriate for galaxies at higher redshifts ($z\gtrsim4$), where the resulting dust masses have been found to greatly exceed what can reasonably be produced by theoretical models \citep{sommovigo2022b,choban2024}. Moreover, at the characteristic redshift of our sample, the CMB temperature reaches $T_\mathrm{CMB}(z=7) = 21.8\,\mathrm{K}$, which means a $25\,\mathrm{K}$ dust reservoir would suffer significant attenuation against the warm CMB \citep{dacunha2013}, and would thus need to be especially massive to have been detected in the current ALMA observations. Finally, recent work by \citet{sommovigo_algera2025} demonstrates that while the dust temperatures obtained from modified blackbody fitting do not necessarily correspond to the true mass- or luminosity-weighted values, they can still be used to reliably infer the dust masses of $z\approx7$ galaxies \citep[see also][]{liang2019,lower2024}.

To account for both the spread and systematic uncertainties in the dust temperatures of our sample, we adopt a fixed error of $\pm15\,$K on the assumed $T_\mathrm{dust}$, and propagate this into our dust mass measurements. We note that upcoming ALMA Band 8 observations of the 8 remaining single-band continuum-detected REBELS-IFU sources will provide more robust dust mass measurements in the future (2024.1.00406.S; PI Algera).

\subsubsection{Molecular gas masses}

Another systematic source of uncertainty is the conversion from \cii{} luminosity to molecular gas mass, for which we adopt $\alpha_\text{\cii{}} 
= 31_{-15}^{+31}\,M_\odot\,L_\odot^{-1}$ as fiducial (Section \ref{sec:gasMassMeasurements}) following \citet{zanella2018}. Different studies, however, have suggested a range of values for $\alpha_\text{\cii{}}$. For example, \citet{madden2020} suggest $\alpha_\text{\cii{}} \approx 72\,M_\odot\,L_\odot^{-1}$ for a sample of local dwarf galaxies (though see \citet{ramambason2024}, who revise their conversion factor downwards to $\alpha_\text{\cii{}} \approx 53\,M_\odot\,L_\odot^{-1}$).\footnote{All quoted values of $\alpha_\text{\cii{}}$ are interpolated/extrapolated to the mean \cii{} luminosity of the REBELS-IFU sample.} \citet{vizgan2022} investigated the \cii{}-to-$M_{\mathrm{H}_2}$ conversion factor by post-processing low-mass $z\approx6$ galaxies drawn from the {\sc{SIMBA}} simulations. For these simulated galaxies, which are comparatively faint in \cii{} ($L_\text{\cii{}} \approx 10^{5 - 7}\,L_\odot$), they find an average $\alpha_\text{\cii{}} \approx 18\,M_\odot\,L_\odot^{-1}$. However, extrapolating their sub-linear relation to the mean \cii{} luminosity of our sample, we infer $\alpha_\text{\cii{}} \approx 87\,M_\odot\,L_\odot^{-1}$. \citet{casavecchia2024} recently determined a relation between molecular gas mass and \cii{} luminosity for the {\sc{ColdSIM}} simulations, for which a similar extrapolation yields $\alpha_\text{\cii{}} \approx 24\,M_\odot\,L_\odot^{-1}$ at the median redshift and \cii{} luminosity of our sample.

We note that a dynamical measurement of $\alpha_\text{\cii{}} = 62_{-40}^{+68}\,M_\odot\,L_\odot^{-1}$ for REBELS-25 by \citet{rowland2024}, based on high-resolution \cii{} imaging, is more or less consistent with all of the prescriptions in the literature. On the other hand, in a study of gravitationally lensed dusty star-forming galaxies at $z\sim4.5$, \citet{rizzo2021} infer a lower average $\alpha_\text{\cii{}} \approx 7\,M_\odot\,L_\odot^{-1}$. Similarly, in a study of $z\approx6$ quasar host galaxies, \citet{kaasinen2024} recently found the \citet{zanella2018} conversion to overestimate CO-based gas masses by a factor of $\sim7$, which thus implies a value of $\alpha_\text{\cii{}} \approx 4.5\,M_\odot\,L_\odot^{-1}$. 

Summarizing, the scatter in literature values for the \cii{}-to-$M_{\mathrm{H}_2}$ conversion factor is thus more than an order of magnitude. We here adopt the \citet{zanella2018} conversion as fiducial not only for consistency with most previous works, but also because it is roughly the average among the aforementioned $\alpha_\text{\cii{}}$ values when expressed in logarithmic units. \\

We parameterize the two key uncertainties -- $T_\mathrm{dust}$ and $\alpha_\text{\cii{}}$ -- and their effect on the dust-to-gas ratio in Fig.\ \ref{fig:grid}. Compared to our fiducial parameter set of $(T_\mathrm{dust}, \alpha_\text{\cii{}}) = (45\,\mathrm{K}, 31\,M_\odot\,L_\odot^{-1})$, the inferred DtG ratio increases by $0.55\,$dex (decreases by $0.3\,$dex) if the true dust temperature is $15\,$K colder (warmer). Similarly, the inferred DtG ratio decreases by $0.5\,$dex (increases by $0.65\,\mathrm{dex}$) if we adopt the extrapolated $\alpha_\text{\cii{}}$ from \citet{vizgan2022} (dynamically inferred value by \citealt{rizzo2021}), compared to our fiducial value. Based on these considerations, and because source-to-source scatter is expected in both the dust temperature and \cii{}-to-H$_2$ conversion factor, we estimate a typical systematic uncertainty of at least $\pm0.5\,$dex on the dust-to-gas and dust-to-metal ratios quoted in the following sections.

\subsubsection{Atomic gas}

On top of the aforementioned uncertainties, the DtG ratio quoted in studies at $z\approx0$ generally includes the contribution from atomic (HI) gas. However, this is typically traced through $21\,\mathrm{cm}$-line observations, which are currently limited to individual unlensed galaxies out to $z\lesssim0.4$ \citep{fernandez2016,xi2024}, or to $z\approx1$ when averaging across galaxy ensembles through stacking (e.g., \citealt{chowdhury2020}). Absorption line spectroscopy against bright background sources, such as quasars or gamma-ray bursts, has further been employed to measure the HI contents of individual foreground galaxies out to $z\sim5$ (see \citealt{peroux2020} for a review). While benefiting from a more or less unbiased selection, such observations typically probe a more extended HI reservoir not well-traced by \cii{} \citep{neeleman2019,heintz2022}. 

Motivated by these difficulties in measuring the atomic gas contents of high-redshift galaxies, \citet{heintz2021} have suggested \ciil{} can additionally be used as a tracer of the HI reservoir in galaxies in the Epoch of Reionization. Applying their metallicity-dependent calibration to our sample, we infer HI gas masses of $M_\mathrm{HI} \approx (0.9 - 6.0) \times M_{\mathrm{H}_2}$, with a median value of $M_\mathrm{HI} \approx 2.1 \times M_{\mathrm{H}_2}$. Taken at face value, this suggests that our dust-to-gas ratios could be overestimated by a typical $\sim0.5\,$dex, due to an underestimation of the total gas mass by an equivalent amount. However, in their chemical and dust evolution framework, \citet{palla2024} suggest that the \citet{heintz2021} calibration may be overestimated at the typically high metallicities of the REBELS sample. Moreover, in their analysis of a metal-poor galaxy at $z=8.50$, \citet{heintz2023} find that the HI mass implied by the \citet{heintz2021} calibration exceeds the dynamical mass of the system, as traced by \cii{}, possibly because the HI reservoir extends to much larger radii. Finally, based on a comparison of \cii{}-based gas masses to the dynamical masses of $z\sim5$ main-sequence galaxies, \citet{dessauges-zavadsky2020} suggest that the gas mass inferred from the \citet{zanella2018} conversion corresponds to the combined molecular and atomic gas masses, as opposed to just the molecular one. This would imply that -- if a substantial atomic gas reservoir is present -- most of its mass is likely found at radii beyond those traced by \cii{}.\footnote{We note the possibility that dust may also extend to larger radii \citep[e.g.,][]{menard2010,peeples2014}. However, as it receives limited radiation from its host galaxy, the dust likely quickly cools rendering it undetectable against the (warm) CMB \citep[see also e.g.,][]{ferrara2025_gsz14}.}

Taken together, these studies thus suggest that the \cii{}-to-HI calibration may not be fully appropriate for (massive) galaxies in the Epoch of Reionization, and/or that the adopted \cii{}-to-H$_2$ conversion may already trace the full gas reservoir. In light of these uncertainties, we do not directly account for atomic gas in the remainder of our analysis. We note that a minor contribution of HI gas is implicitly accounted for by our previously quoted systematic error of $\sim0.5\,$dex. In the future, high-resolution \cii{} observations of REBELS galaxies will enable dynamical constraints on their atomic and molecular gas masses (c.f., the existing dynamical constraints by \citealt{rowland2024} for REBELS-25; see also Phillips et al.\ in preparation).

\begin{figure*}
    \centering
    \includegraphics[trim={0 0.7cm 0 0}, width=0.7\textwidth]{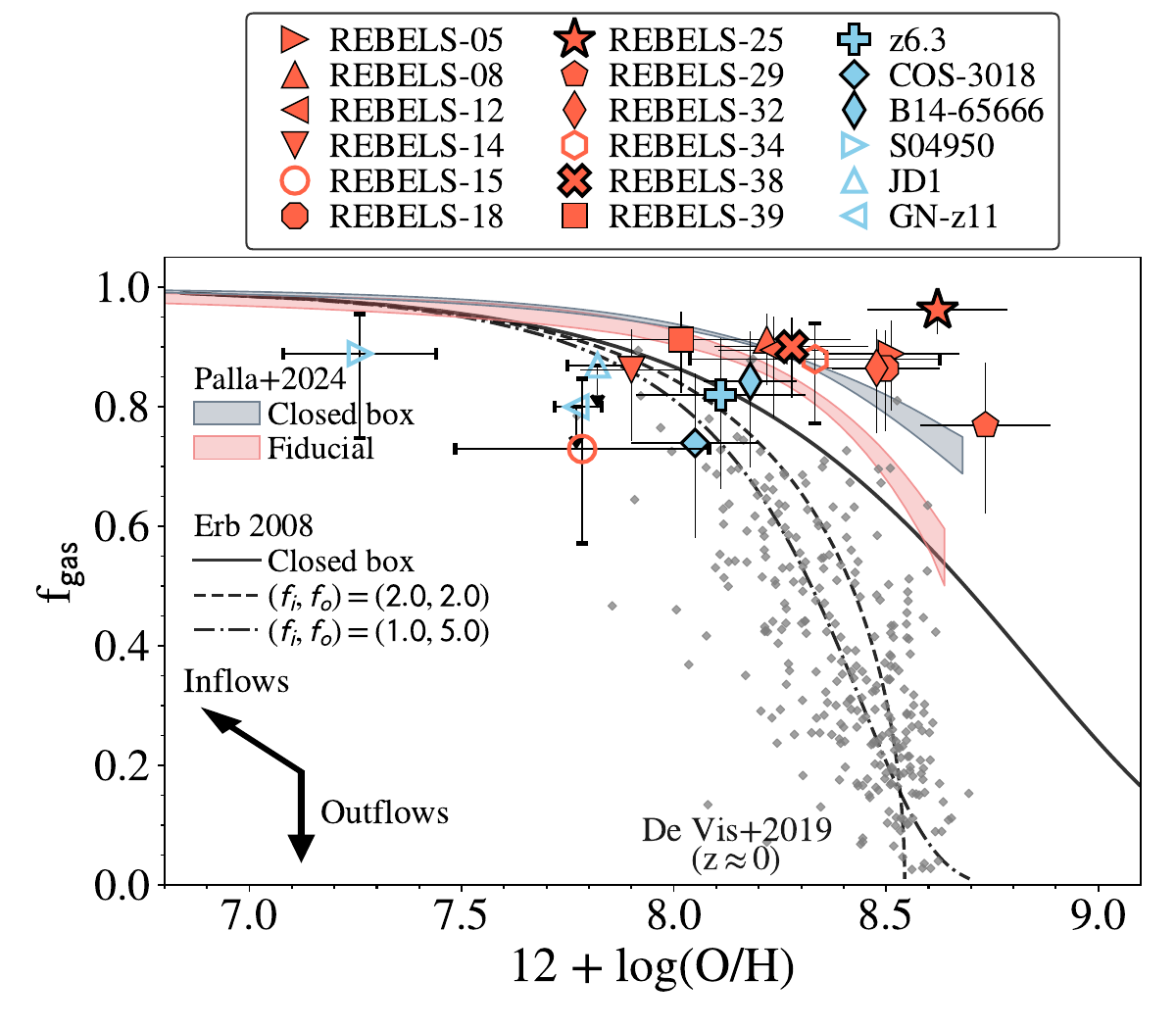}
    \caption{Gas fraction as a function of metallicity for the REBELS-IFU sample (orange), six $z>6$ literature galaxies with \cii{} observations and metallicity measurements (light blue; Appendix \ref{app:literature}) and local galaxies from \citet{devis2019}. We overlay the model tracks from \citet{palla2024}, both using their fiducial model, and a closed-box model. The latter, lacking in- or outflows, provides the theoretical maximum gas fraction for a given metallicity. The width of the model tracks represents the variation in $f_\mathrm{gas}$ based on the adopted star-formation history. Moreover, we overlay tracks from the analytical model by \citet{erb2008}, whereby $f_i$ ($f_0$) represents the inflow (outflow) rate in units of the SFR. The effect of in- and outflows on the gas fraction and metallicity is shown schematically by the arrows in the lower left corner, under the simplifying assumption that outflows deplete molecular gas and metals equally. At low metallicities [$12+\log(\mathrm{O/H})\sim 7.8 \sim 0.1\,Z_\odot$], the REBELS and literature samples show lower gas fractions than the model predictions, while at high metallicities [$12+\log(\mathrm{O/H})\gtrsim 8.5 \sim 0.6\,Z_\odot$] their gas fractions match or even exceed the closed box model.}
    \label{fig:gasFractionMetallicity}
\end{figure*}

\section{Results}
\label{sec:results}

\subsection{Gas fractions}
\label{sec:FgasMetallicity}

We start by investigating the gas fractions [$f_\mathrm{gas} = M_{\mathrm{H}_2} / (M_{\mathrm{H}_2} + M_\star$)] of our sample as a function of metallicity in Fig. \ref{fig:gasFractionMetallicity}. This comparison serves two purposes: first of all, galaxies are expected to trace out an evolutionary pathway in this parameter space, with the formal starting point being a gas fraction of unity and zero metallicity (e.g., \citealt{tinsley1980,finlator2008}). As they evolve, galaxies move along a path shaped by star formation -- which reduces their gas fractions and increases their metallicities -- as well as by (primordial) gas inflows and (metal-enriched) outflows. For a given metallicity, the gas fraction is maximized in a simple model without in- and outflows known as the \textit{closed box model}. In addition to providing insight into the evolutionary pathways of our sample, the $f_\mathrm{gas}$ - metallicity diagram also provides a useful practical check on the accuracy of our gas mass and stellar mass measurements, as a combination of both a high $f_\mathrm{gas}$ and high metallicity is unphysical -- this would suggest that the gas mass or metallicity is overestimated, the stellar mass is underestimated, or a combination of the above.

Figure \ref{fig:gasFractionMetallicity} shows that the REBELS-IFU galaxies are characterized by high gas fractions of $f_\mathrm{gas} \approx 0.73 - 0.96$, with a mean value of $\langle f_\mathrm{gas} \rangle = 0.87 \pm 0.06$ where the error represents the standard deviation among the sample. These values are broadly consistent with the handful of $z\gtrsim6$ galaxies with \cii{} and metallicity measurements in the literature (see Appendix \ref{app:literature} for details), although the REBELS-IFU sample is on average more metal-enriched.

We compare the gas fractions of the REBELS-IFU and literature galaxies to the chemical evolutionary tracks from \citet{palla2024}, who use the non-parametric SFHs derived for three REBELS galaxies (REBELS-15, REBELS-25 and REBELS-29) as an input to investigate their dust and metal build-up (labeled `fiducial' in Figure \ref{fig:gasFractionMetallicity}).\footnote{Note that these SFHs, taken from \citet{topping2022}, predate the \textit{JWST} observations of the REBELS sample. Updated SFHs of the full REBELS-IFU sample will be presented in Stefanon et al.\ (in preparation).} Moreover, we include a re-run of the \citet{palla2024} models without in- and outflows, i.e., a closed box model. 

We furthermore compare to the \citet{erb2008} evolutionary tracks in Figure \ref{fig:gasFractionMetallicity}. Their simple analytical prescription adopts inflow and outflows rates that are proportional to the SFR by a factor $f_i$ and $f_o$, respectively. Comparing the \citet{palla2024} and \citet{erb2008} closed box models, the latter predict lower gas fractions at a fixed metallicity. This is primarily because of the different stellar yields adopted: while \citet{erb2008} adopt a fixed metal yield of $y=0.02$ per unit of formed stellar mass, \citet{palla2024} use yields specific for low-intermediate mass stars and core-collapse SNe (see also \citealt{spitoni2017}). 

Regardless of the precise assumptions regarding stellar yields and outflows, the \citet{erb2008} and \citet{palla2024} models predict similar global trends: unsurprisingly, as galaxies become more enriched, their gas fractions decrease. Interestingly, this prediction is not mirrored by the REBELS-IFU and literature samples. At low metallicities, both REBELS-15 and several literature galaxies such as COS-3018, JD1 and GN-z11 appear to have low gas fractions, while some of the most metal-rich sources seemingly have a high $f_\mathrm{gas}$. In particular, some of the most metal-rich REBELS-IFU galaxies have gas fractions in excess of even the closed box model predictions, which should constitute a stringent upper limit. Barring some contrived scenario involving metal-rich inflows and/or metal-poor outflows, this suggests that the gas masses and/or metallicities of (some of) the REBELS-IFU sources may be overestimated, or that their stellar masses are underestimated. We discuss this in the following subsections.

\subsubsection{Underestimated stellar masses?}

One potential avenue of resolving the seemingly unphysical combination of high metallicities and gas fractions, is if the stellar masses of the REBELS-IFU sources are underestimated (i.e., their gas fractions are overestimated). However, for only one of our targets do we have evidence that this may be the case: the source with $f_\mathrm{gas}$ most strongly in excess of the closed box models is the well-studied $z=7.31$ galaxy REBELS-25 (\citealt{hygate2023,algera2024b,rowland2024}). This source benefits from a dynamical mass measurement (\citealt{rowland2024}), and the dynamically inferred $\alpha_\text{\cii{}} \sim 60\,M_\odot\,L_\odot^{-1}$ suggests the gas mass is unlikely to be significantly overestimated using the \citet{zanella2018} prescription, unless REBELS-25 is strongly dark-matter-dominated. Instead, several pieces of evidence suggest its stellar mass may be underestimated. First of all, REBELS-25 shows a peculiar morphology, with an IR-luminous, centrally-peaked dust distribution, surrounded by three UV-luminous clumps (\citealt{hygate2023,rowland2024}). This makes it plausible that its stellar mass is underestimated due to outshining by these UV-bright regions (c.f., \citealt{gimenez-arteaga2024}). An underestimated stellar mass is independently supported by the high dust mass measured for REBELS-25 by \citet{algera2024b}, which suggests a dust-to-stellar mass ratio of $\sim6\,\%$, in strong tension with models of dust production (see the discussion in \citealt{algera2024b}, and Section \ref{sec:discussionComparisonSimulations}).

On the other hand, none of the other REBELS-IFU sources show such a striking offset between their UV and dust emission, suggesting that REBELS-25 may be unique in this regard (although high-resolution ALMA observations of the full sample are required to verify this). In addition, recent works at slightly lower redshift ($z\sim5$) have shown that the effects of outshining may be limited at relatively high stellar masses [$\log(M_\star/M_\odot) \gtrsim 9$], based on a comparison of spatially-resolved and integrated SED modeling \citep{li2024,lines2024}. As these studies span a similar mass range as the galaxies in our work, this could suggest outshining not to be a major issue for the REBELS-IFU sample as a whole. However, we stress that the exact magnitude of outshining in resolved versus integrated photometry remains a topic of debate in the high-redshift literature \citep[e.g.,][]{mosleh2025,harvey2025}. For this reason, Laza-Ramos et al.\ (in preparation) carry out spatially-resolved SED-fitting on the NIRSpec/IFU observations of the twelve targets presented in this work. Their analysis suggests that, while resolved stellar masses indeed slightly exceed those from integrated photometry, the overall effect is minor, with an average offset of $<0.3\,\mathrm{dex}$. This modest increase in the average stellar masses of our sample does not significantly affect our conclusions.

In addition to the above, the magnitude of outshining can also depend on the assumed star formation history (e.g., \citealt{leja2017,leja2019}). For example, \citet{topping2022} compare the stellar masses obtained for the REBELS sample by fitting constant and non-parametric SFHs to their integrated (pre-\textit{JWST}) photometry. They find that the non-parametric SFH yields consistently higher masses, by an average of $\sim0.4\,\mathrm{dex}$ (see also e.g., \citealt{narayanan2024_outshining,cochrane2025} for recent work on this topic). In our analysis, however, we already rely on non-parametric stellar masses in an attempt to circumvent this issue. This makes it unlikely that our stellar masses are currently significantly underestimated due to assumptions regarding the star formation history, although we note the possibility that deeper IFU observations could reveal a faint, older stellar population beyond the current surface brightness limits of our data.

We note that -- in addition to the aforementioned spatially-resolved SED-fitting -- future observations using \textit{JWST}/MIRI would also enable more accurate stellar mass determinations for our targets (c.f., \citealt{wang2024}). For now, however, we proceed with the fiducial stellar mass measurements from Stefanon et al.\ (in preparation), as presented in Table \ref{tab:data} and also used in \citet{rowland2025}. While the potential systematic uncertainties on these measurements should be borne in mind, we emphasize that the core of our analysis focuses on the dust-to-gas and dust-to-metal ratios of the REBELS-IFU sample, and hence does not rely on stellar mass estimates.

\subsubsection{Overestimated metallicities?}

The high $f_\mathrm{gas}$ of the metal-rich REBELS-IFU sources could also be the result of their metallicities being overestimated. Indeed, \citet{rowland2025} find that the average metallicity of the REBELS-IFU galaxies decreases slightly when the $O32$ diagnostic is adopted for the full sample (see also Appendix \ref{app:metallicity}). Such lower metallicities would reduce the tension with the models somewhat, although would not resolve it completely. Moreover, it would not significantly affect the galaxies with low fiducial metallicity measurements; these would still appear gas-poor with respect to the \citet{palla2024} models. Observations of auroral lines such as $\mathrm{[O\,III]}_{4363}$ (e.g., \citealt{curti2023,heintz2023}) would provide more robust metallicity estimates, and would enable investigating whether the current strong-line metallicities are indeed being overestimated.

\subsubsection{Overestimated gas masses?}

Finally, it is possible -- and perhaps most likely -- that the gas masses of the REBELS-IFU sample are overestimated. We extensively discussed the effect of $\alpha_\text{\cii{}}$ on our analysis in Section \ref{sec:errorBudget}, and recall that it shows over an order of magnitude of scatter in the literature. One way of reconciling the REBELS-IFU + literature sources with the expected $f_\mathrm{gas}$-metallicity relation, is to suppose $\alpha_\text{\cii{}}$ varies with metallicity. This would not be too far-fetched: studies of local dwarf galaxies find relatively large values of $\alpha_\text{\cii{}}$ (e.g., \citealt{madden2020,ramambason2024}), although within their limited dynamic range they do not find any clear trends with metallicity. On the other hand, dynamical studies of dusty star-forming galaxies and high-$z$ quasars -- likely metal-rich populations -- find significantly lower values (e.g., \citealt{rizzo2021,kaasinen2024}). However, direct observational evidence for a metallicity-dependent $\alpha_\text{\cii{}}$ is currently lacking (see also \citealt{zanella2018}).

Among theoretical studies, there is similarly no clear consensus on whether $\alpha_\text{\cii{}}$ varies with metallicity. For instance, \citet{vizgan2022} and \citet{gurman2024} find no or only a weak metallicity-dependence, while recent works by \citet{casavecchia2025} and \citet{vallini2025} find that $\alpha_\text{\cii{}}$ increases for metal-poor galaxies. Within the metallicity range spanned by the REBELS-IFU sample, however, the variation in $\alpha_\text{\cii{}}$ predicted by these studies is at maximum a factor of two. We therefore proceed with our fiducial gas mass measurements assuming the fixed \citet{zanella2018} conversion for the entire sample.

We emphasize that if the gas masses of our metal-rich sample are currently overestimated, our dust-to-gas and dust-to-metal ratios would instead be underestimated; higher DtG and DtM ratios would further reinforce our conclusions in Section \ref{sec:discussion} that dust build-up in the REBELS-IFU sample proceeds in a highly efficient manner.

With these caveats mind, we prefer not to over-interpret the possibly inverted trend between $f_\mathrm{gas}$ and metallicity with respect to the models. We therefore proceed with our fiducial stellar and gas masses and metallicities, and use these to investigate key scaling relations between dust, gas and metals in the next section.

\subsection{Dust scaling relations}
\begin{figure*}
    \centering
    \includegraphics[trim={0 0.5cm 0 0},width=0.9\textwidth]{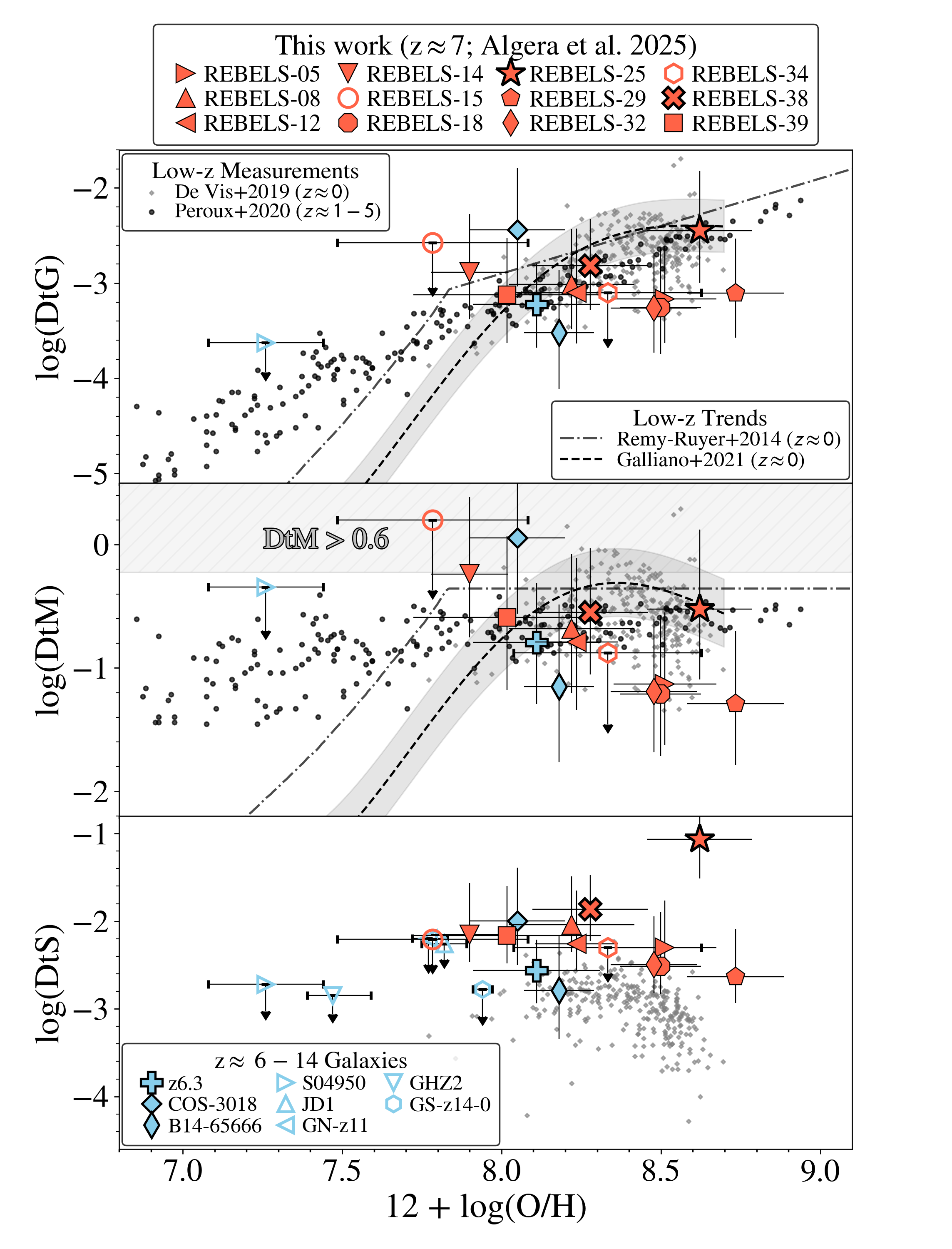}
    \caption{The dust-to-gas (DtG; \textit{top panel}), dust-to-metal (DtM; \textit{middle}) and dust-to-stellar mass (DtS; \textit{bottom}) ratios as a function of gas-phase metallicity, drawn from a variety of low-, intermediate- and high-redshift observational studies. The REBELS-IFU sources, as well as additional high-redshift galaxies with constraints on their dust and metal contents (Appendix \ref{app:literature}), are shown in orange and light blue, respectively. Galaxies with filled markers are dust-detected in either a single band (narrow black outline) or multiple bands (thick black outline and larger marker). Open markers represent galaxies without a dust continuum detection. At low and intermediate redshifts, we show either individual galaxies \citep{devis2019,peroux2020}, or average trends \citep{remyruyer2014,galliano2021}. The REBELS-IFU galaxies, at relatively high metallicities of $Z\approx0.1-1.1\,Z_\odot$, show similar or slightly lower DtG and DtM ratios compared to local galaxies. Conversely, their DtS ratios are larger than those for galaxies at $z\approx0$.}
    \label{fig:DustBuildupObservations}
\end{figure*}

We consider three key scaling relations as a function of metallicity: the dust-to-gas, dust-to-metal and dust-to-stellar mass (DtS) ratios (Figures \ref{fig:DustBuildupObservations} and \ref{fig:DustBuildupSimulations}). Figure \ref{fig:DustBuildupObservations} includes a comparison of the REBELS-IFU sample to observations at low ($z\approx0$), intermediate ($z\approx1-5$) and high ($z\gtrsim6$) redshifts, while Figure \ref{fig:DustBuildupSimulations} includes predictions from a set of theoretical models at $z\approx7$. We discuss these comparisons to observations and simulations in detail in Section \ref{sec:discussion}, while we first focus on some of the general properties of the REBELS-IFU sample. 

The top panels of Figures \ref{fig:DustBuildupObservations} and \ref{fig:DustBuildupSimulations} demonstrate that the REBELS-IFU sources span nearly $1\,$dex in DtG ratio, ranging from $-3.3 \lesssim \log(\mathrm{DtG}) \lesssim -2.4$. However, the dust-to-gas ratio does not appear to vary strongly with metallicity above $12 + \log(\mathrm{O/H}) \gtrsim 8$, although for our lowest-metallicity target we can only provide an upper limit on its dust mass, and thus on the DtG ratio. The source with the largest DtG ratio is REBELS-25 at $z=7.31$, which is simultaneously the galaxy with the highest dust mass among the REBELS sample \citep{algera2024b}. Among the 10 dust-detected galaxies, we find a mean dust-to-gas ratio of $\langle \log(\mathrm{DtG}) \rangle = -3.02 \pm 0.23$, where the uncertainty represents the standard deviation among the sample and does not include any systematic errors on the dust and gas masses.

In the middle panels of Figures \ref{fig:DustBuildupObservations} and \ref{fig:DustBuildupSimulations}, we plot the dust-to-metal ratio against metallicity. Because of the flat trend in dust-to-gas ratio versus metallicity, the DtM ratio shows an apparent decrease with $12+\log(\mathrm{O/H})$ (we recall that $\mathrm{DtM} = \mathrm{DtG} / Z$). We demarcate the line with the expected maximum possible dust-to-metal ratio of $\mathrm{DtM} = 0.6$ (\citealt{palla2024}; see also \citealt{konstantopoulou2024}). Encouragingly, our entire sample is consistent with this strict upper limit. Focusing once more on the 10 dust-detected galaxies, we calculate the \textit{total} fraction of metals locked up in dust as $\mathrm{DtM}' = M_\mathrm{dust} / (M_\mathrm{dust} + M_Z) = \mathrm{DtM} / (1 + \mathrm{DtM})$. Our sample spans a range of $\mathrm{DtM}' = 0.05 - 0.36$, with an average value of $\mathrm{DtM}' = 0.16 \pm 0.10$. This highlights how a substantial fraction of metals can be locked up in dust, even in galaxies at $z\approx7$. 

Finally, the bottom panels of Figures \ref{fig:DustBuildupObservations} and \ref{fig:DustBuildupSimulations} plot the ratio of dust-to-stellar mass against metallicity. With the exception of REBELS-25, whose stellar mass may be underestimated (Sec.\ \ref{sec:FgasMetallicity}), the REBELS-IFU sample spans a range of $-2.7 \lesssim \log(\mathrm{DtS}) \lesssim -1.8$, in agreement with a previous investigation of this quantity by \citet{dayal2022} (see also \citealt{sommovigo2022}). The DtS ratio does not show a clear trend with metallicity, at least for the relatively metal-rich REBELS-IFU sample with $12+\log(\mathrm{O/H}) \gtrsim 8.0$. We tabulate the DtG, DtM and DtS ratios of our targets in Table \ref{tab:data}.

\section{Discussion}
\label{sec:discussion}

\subsection{A comparison to observations}
\label{sec:discussionComparisonObservations}

In Figure \ref{fig:DustBuildupObservations}, we compare the DtG, DtM and DtS ratios of the REBELS sample to a broad range of observations spanning $z\approx0$ to $z\approx14$. We start by briefly introducing these comparison samples. 

In the local Universe, we compare to observations from \citet{remyruyer2014}, \citet{devis2019} and \citet{galliano2021}. We note that their samples partially overlap, and are therefore not fully independent. Each of the local studies find an increase in the DtG ratio towards more metal-rich galaxies, with both \citet{remyruyer2014} and \citet{galliano2021} preferring a rapid increase with metallicity for $12 + \log(\mathrm{O/H}) \lesssim 8$, while above this turnover metallicity the increase in DtG ratio becomes approximately linear. As a result, in these studies the DtM ratio does not depend strongly on metallicity above the turnover.

Some of the first direct constraints on the DtG ratios of galaxies at higher redshift ($z\sim1.5-2.5$) were obtained by \citet{shapley2020}, who presented Keck/MOSFIRE follow-up of four ALMA-detected galaxies from the ASPECS program \citep{walter2016} to measure their rest-optical metallicities. Both their study and later work by \citet{popping2023}, who studied the dust and CO emission in 10 galaxies with rest-optical spectra drawn from the MOSDEF survey (\citealt{kriek2015}; see also \citealt{shivaei2022}), found good agreement with the local relation between DtG ratio and metallicity. This suggests the local scaling relation to already be in place at Cosmic Noon -- at least at the high-metallicity end to which the aforementioned studies are limited [$12 + \log(\mathrm{O/H}) \gtrsim 8.4$].

Also at intermediate redshifts ($z\approx 1 - 5$), \citet{peroux2020} have compiled DtG and DtM measurements for DLAs. These samples appear to show a linear trend between DtG and metallicity across the full parameter space, without clear evidence for a turnover at intermediate metallicities. While the reason for this difference with respect to $z\approx0$ galaxies is not fully clear, the discrepancy could be related to systematic differences between studies using absorption- and emission-line spectroscopy (\citealt{hamanowicz2020,hamanowicz2024,clark2023,park2024}).

At even earlier epochs, combined studies of dust and metals are scarce. We here compare to eight $z > 6$ galaxies with dust and metal observations in the literature, six of which were moreover targeted in \cii{} emission enabling insight into their gas properties. Details of our literature sample are provided in Appendix \ref{app:literature}. Briefly, however, our sample consists of RXCJ0600-z6.3 at $z=6.07$ (a.k.a. the `Cosmic Grapes'; \citealt{fujimoto2024b,gimenez-arteaga2024,valentino2024}); COS-3018 at $z=6.85$ \citep{smit2018,witstok2022,scholtz2024}; B14-65666 at $z=7.15$ \citep{hashimoto2019,sugahara2021,jones2024}; S04950 at $z=8.50$ \citep{heintz2023,fujimoto2024a}; JD1 at $z=9.11$ \citep{hashimoto2018,tokuoka2022,marconcini2024,morishita2024}; GN-z11 at $z=10.60$ \citep{oesch2016,bunker2023,tacchella2023,fudamoto2024}; GHZ2 at $z=12.33$ \citep{castellano2022,castellano2024,naidu2022,bakx2023,calabro2024,zavala2024,zavala2024b}; and GS-z14-0 at $z=14.18$ \citep{carniani2024,carniani2024b,schouws2024}. The $6.07 \leq z \leq 7.15$ sample are all detected in \cii{} and dust continuum, while S04950 is \cii{}- but not dust-detected. The \cii{} lines in JD1 and GN-z11 were targeted but not detected, and neither was their dust continuum; for those sources we can therefore not place any meaningful constraints on their DtG and DtM ratios. Finally, GHZ2 and GS-z14-0 have not been targeted in \cii{} emission, although for the latter ALMA Director's Discretionary Time (DDT) was recently awarded to observe the line (2024.A.00007.S, PI Schouws).\footnote{While this manuscript was in the refereeing stage, the results of these observations were presented in \citet{schouws2025}, with \cii{} remaining undetected in GS-z14-0.}

\subsubsection{Dust-to-Gas and Dust-to-Metal ratios}
In the top and middle panels of Figure \ref{fig:DustBuildupObservations}, we see that the DtG and DtM ratios of the REBELS-IFU sources are broadly consistent with those of local galaxies at similar gas-phase metallicities. While we do not see evidence for a linear trend between DtG and metallicity -- unlike the local studies -- this could simply be due to the fixed dust temperature and $\alpha_\text{\cii{}}$ we adopt to obtain the dust and gas masses. This appears plausible when considering only the $z \gtrsim 6$ galaxies with multi-band dust mass and metallicity measurements (points with a thick black outline in Figure \ref{fig:DustBuildupObservations}, which are in order of increasing redshift: z6.3, REBELS-38, COS-3018, B14-65666 and REBELS-25). Between just these five sources, there does appear to be an increasing trend of DtG with metallicity.

As described in Section \ref{sec:errorBudget}, we adopt a fixed dust temperature of $T_\mathrm{dust} = 45 \pm 15\,\mathrm{K}$ for the remainder of the REBELS-IFU sample to infer their dust masses. However, the theoretical model by \citet{sommovigo2022} suggests that dust temperatures may be lower for metal-rich galaxies, which would consequently increase their dust mass and DtG ratio compared to our fiducial values. Indeed, as discussed in Section \ref{sec:errorBudget}, if the metal-rich sample has a lower dust temperature of $T_\mathrm{dust} \approx 30 - 35\,\mathrm{K}$, their DtG and DtM ratios increase by $\sim0.5\,\mathrm{dex}$, bringing them into better agreement with the low-redshift relations. On the other hand, for local galaxies there does not appear to be a clear relation between dust temperature and gas-phase metallicity (e.g., \citealt{lamperti2019}). Recently-approved ALMA Band 8 observations of the single-band continuum-detected REBELS-IFU galaxies will shed light on this by enabling dual-band dust temperature and mass measurements (2024.1.00406.S; PI Algera).

Consistent with the flat relation between DtG and metallicity, we see a slight decreasing trend between DtM and $12 + \log(\mathrm{O/H})$ (middle panel of Fig.\ \ref{fig:DustBuildupObservations}). This trend is not seen in lower redshift galaxies, and may again be the result of uncertain dust and gas mass measurements. However, if taken at face value, there are a few possible explanations for this low DtM at high metallicity. First of all, observations at low-redshift suggest that dust growth in the ISM should be especially efficient at high metallicities (e.g., \citealt{mattsson2014,galliano2021,konstantopoulou2024}). However, the characteristic timescale of dust growth is poorly constrained, and likely depends on the physical conditions within the galaxy. If the timescale is sufficiently long, a galaxy following a recent starburst may have enriched its ISM with metals, without having yet locked these metals into significant amounts of dust. \citet{galliano2021} attempt to empirically constrain the timescale for dust growth in local galaxies, finding it to be between $0.1 - 1\,\mathrm{Gyr}$. Even at the lower end of this range, this is comparable to the typical ages of high-redshift galaxies (e.g., \citealt{whitler2023}).

On the other hand, given the higher characteristic ISM densities in the early Universe, it is plausible that dust growth timescales are significantly shorter in distant galaxies. \citet{algera2024b} argued this is likely the case for the $z=7.31$ galaxy REBELS-25, which possesses a particularly high dust mass and moreover shows a steep dust emissivity index that could point towards a dense ISM where dust growth is particularly efficient (see also e.g., \citealt{stepnik2003,kohler2015}). Clearly, more detailed modeling of ISM grain growth physics in dense or otherwise extreme physical conditions is necessary to test this scenario.

Another possible explanation is that dust destruction becomes increasingly dominant in metal-rich galaxies at high redshift. Indeed, some models have suggested that SNe may be net dust destroyers when embedded in a dust-rich ISM (e.g., \citealt{bocchio2016,kirchschlager2024}). Assuming that each SN is capable of sweeping up a fixed ISM mass, the total swept up dust mass will be proportional to the dust-to-gas ratio (e.g., \citealt{asano2013,dayal2022}), which would imply SN dust destruction to become increasingly important in more dust-rich galaxies. Given that these SNe would simultaneously release metals previously locked up in dust grains back into the gas phase, a sudden increase in SN dust destruction could potentially produce a population of metal-rich yet dust-poor galaxies. In the local Universe, a slight decrease in galaxy dust masses and DtM ratios is indeed seen at the high metallicity end (e.g., \citealt{devis2019,galliano2021}), which is due to older, quiescent galaxies where little dust is expected to be produced, and dust destruction becomes dominant (see also \citealt{calura2023,donevski2023,lee2024}). However, the rest-optical spectra of the REBELS-IFU sources demonstrate these are all clearly star-forming galaxies, where this behavior is not expected. Indeed, none of the models shown in Figure \ref{fig:DustBuildupSimulations} (and discussed in Section \ref{sec:discussionComparisonSimulations}) predict a trend of this nature, despite otherwise being successful in reproducing a wide range of properties of the high-redshift galaxy population. As such, we disfavor this interpretation. 

Instead of completely destroying the dust, it is also possible that dust is effectively ejected from the galaxy (e.g., \citealt{ferrara2023,otsuki2024}). However, if dust and gas are well-mixed, outflows should leave the DtG and DtM ratios unchanged, and as such this interpretation requires the preferential ejection of dust. One possibility is that the dust is ejected due to strong radiation pressure (e.g., \citealt{ferrara2023,ziparo2023}). However, as shown by \citet{ferrara2024_lya}, this scenario requires highly compact galaxies with large specific star formation rates (sSFRs), whereas the REBELS-IFU galaxies are both spatially extended \citep{rowland2025}, and have lower sSFRs (\citealt{topping2022}; Fisher et al.\ in preparation). Moreover, the outflow interpretation is disfavored by the high gas fractions inferred for the REBELS-IFU sample (Section \ref{sec:FgasMetallicity}). Similarly, while inflows of primordial gas could be invoked to explain a reduction in the DtG ratio, these cannot reduce the DtM ratio. This suggests that, if the lower DtM ratios at high metallicities are real, efficient dust destruction or inefficient ISM dust growth are more plausible explanations.

For now, however, we take this trend as rather tentative -- the metal-rich galaxies showing low DtM ratios are the same ones showing gas fractions in excess of the closed box model (Section \ref{sec:FgasMetallicity}). As such, we deem it more likely that their gas masses are currently being overestimated, which would bring their DtG and DtM ratios in better agreement with the local relations.

\subsubsection{Dust-to-Stellar mass ratio}

The lower panel of Figure \ref{fig:DustBuildupObservations} reveals that the dust-to-stellar mass ratios of the REBELS-IFU sample span a range of $-2.7 \lesssim \log(\mathrm{DtS}) \lesssim -1.8$, after excluding REBELS-25 with $\log(\mathrm{DtS}) \approx -1.1$ for which the stellar mass is likely underestimated (Section \ref{sec:FgasMetallicity}). These DtS ratios are in agreement with previous determinations for the REBELS sample adopting the \cii{}-based dust temperature estimated by \citet{sommovigo2022}. Moreover, the inferred DtS ratios are similar to those observed in the typical $z\gtrsim4$ galaxy population \citep{witstok2023}, and are furthermore similar to the what is observed in dusty star-forming galaxies at cosmic noon (e.g., \citealt{dacunha2015,donevski2020,dudzeviciute2020,liao2024}). However, the DtS ratios measured for the REBELS-IFU galaxies are larger than those of the local sample from \citet{devis2019}, which includes a mixture of dwarf, spiral and elliptical galaxies (see also \citealt{davies2017}). 

The fact that the DtS ratio does not appear to scale with metallicity within the REBELS-IFU sample could suggest that the dust is produced solely through Type II SNe (though see the discussion in the following Section). Following the framework by \citet{michalowski2015}, we convert the measured DtS ratios of the REBELS-IFU sample to a Type-II SN yield assuming a \citet{chabrier2003} IMF. Focusing on the dust-detected galaxies and excluding REBELS-25 due to its likely underestimated stellar mass, we infer an average yield of $y_\mathrm{SN} = 0.52 \pm 0.28\,M_\odot\,\mathrm{SN}^{-1}$. This is consistent with the values previously reported for the continuum-detected REBELS sample, which partially overlaps with the REBELS-IFU sample \citep{ferrara2022,sommovigo2022}, as well as for literature samples at $z\gtrsim4$ (e.g., \citealt{witstok2023}).  

However, we recall that the dust continuum detections among the REBELS-IFU sample are limited to relatively high metallicities [$12 + \log(\mathrm{O/H}) \gtrsim 8$, corresponding to $Z\gtrsim0.2\,Z_\odot$]. Several studies have suggested that this exceeds a `critical metallicity' threshold where dust growth in the ISM starts to become particularly efficient (e.g., \citealt{asano2013,zhukovska2014,parente2022,choban2024a}). Indeed, at metallicities far below this expected threshold, the low DtS ratio of S04950 [$\log(\mathrm{DtS}) \lesssim -2.8$] could point towards lower supernova yields. The DtS ratios of GHZ2 and GS-z14-0 are similarly low, and these galaxies are found at an epoch ($z\gtrsim12$) where dust growth in the ISM is less likely to significantly contribute to overall dust build-up. Based on these three sources alone, the inferred SN yield is therefore significantly lower, $y_\mathrm{SN} < 0.1 - 0.15\,M_\odot\,\mathrm{SN}^{-1}$ given a Chabrier IMF. 

We caution, however, that the dust masses of these sources are upper limits based on ALMA non-detections, which require assuming a dust temperature. As detailed in Section \ref{sec:errorBudget}, this can significantly affect inferred dust masses. Moreover, the stellar masses of these high-redshift galaxies are similarly uncertain (e.g., \citealt{helton2024}). In addition, converting the DtS ratio into a supernova yield implicitly assumes that all produced dust still resides within the galaxy, whereas for these ultra-high-redshift sources dust may be efficiently expelled \citep{ferrara2023,ferrara2024}. This would cause the observationally-inferred dust mass to underestimate the total amount of dust the galaxy has produced across its lifetime, and thus result in underestimated SN yields. As the above discussion suggests, and as we will see in the following Section, it is (unfortunately) not straightforward to distinguish between supernova and ISM dust production in the early Universe.

\subsection{A comparison to models}
\label{sec:discussionComparisonSimulations}

\begin{figure*}
    \centering
    \includegraphics[trim={0 0.5cm 0 0}, width=0.9\textwidth]{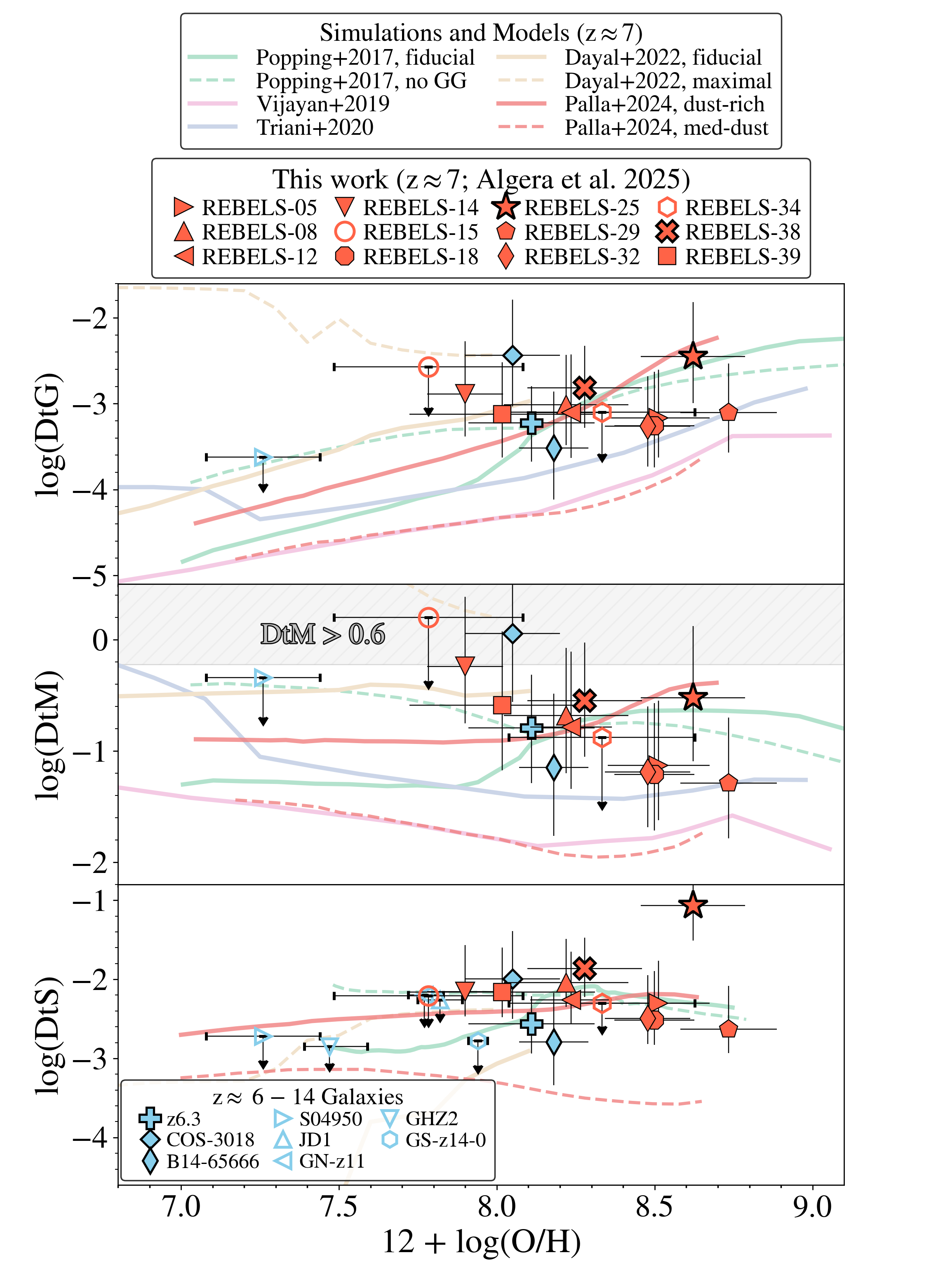}
    \caption{The dust-to-gas (\textit{upper}), dust-to-metal (\textit{middle}) and dust-to-stellar mass (\textit{bottom}) ratios as a function of gas-phase metallicity. Similar to Fig.\ \ref{fig:DustBuildupObservations}, we overplot the twelve REBELS-IFU sources and other $z\gtrsim6$ galaxies in the literature. Moreover, we compare to a variety of studies modeling the production and evolution of dust at $z\approx7$ \citep{popping2017,vijayan2019,triani2020,dayal2022,palla2024}. Overall, our measurements are best reproduced by models predicting rapid dust build-up, resulting in high DtG, DtM and DtS ratios across the metallicity range spanned by the REBELS-IFU sample.}
    \label{fig:DustBuildupSimulations}
\end{figure*}

To gain further insight into the dust production and destruction pathways of the REBELS-IFU sample, we proceed by comparing to several models of early dust build-up (Fig.\ \ref{fig:DustBuildupSimulations}). Specifically, we compare to two different prescriptions by \citet{popping2017}, who include dust production and destruction on top of the semi-analytical Santa Cruz models: we focus on their fiducial model, and an alternative prescription that does not include ISM dust growth (labeled `no GG' in Fig.\ \ref{fig:DustBuildupSimulations}). Moreover, we compare to the L-Galaxies semi-analytical models from \citet{vijayan2019}, the DustySAGE models introduced by \citet{triani2020}, and the DELPHI models by \citet{dayal2022}. The latter study considers two scenarios: a fiducial model whereby most dust is created by SNe, and a maximal model with rapid ISM dust growth and no dust destruction, which ensures that effectively all metals are found inside dust grains. Finally, we consider the models by \citet{palla2024}, which couple the observationally-constrained star formation histories of REBELS galaxies from \citet{topping2022} to a chemical evolutionary framework. We consider both the metal-rich/dust-rich and metal-rich/intermediate-dust (labeled `med-dust' in Fig.\ \ref{fig:DustBuildupSimulations}) scenarios described by \citet{palla2024}. For each of the models, we focus on the predictions made at $z\approx7$. While not an exhaustive list of works investigating early dust build-up (see also, e.g., the recent studies by \citealt{dicesare2023,lewis2023,choban2024,esmerian2024,narayanan2024}), the selected comparison studies span a large range of predictions in terms of the DtG and DtM ratios, and thus provide insight into what may be the dominant processes governing the production and evolution of dust at $z\approx7$. Where relevant, we moreover include a qualitative comparison to other works not shown in Fig.\ \ref{fig:DustBuildupSimulations}. 

\subsubsection{Dust-to-Gas and Dust-to-Metal ratios}

In terms of the DtG and DtM ratios, the majority of the REBELS-IFU sample is in good agreement with the dust-rich scenario in \citet{palla2024}, or either of the two dust production scenarios from \citet{popping2017}. Both of these studies predict high dust-to-gas ratios [$\log(\mathrm{DtG}) \sim -3$] at the metallicity range of the REBELS-IFU sample [$12 + \log(\mathrm{O/H}) \gtrsim 8$]. In the case of the fiducial model by \citet{popping2017}, the large DtG and DtM ratios are predominantly due to dust growth in the ISM, although their `no GG' model suggests efficient supernovae can produce similarly high values. Constraints on the DtG and DtM ratios for galaxies at lower metallicities are essential to distinguish between these two scenarios; at metallicities below $12 + \log(\mathrm{O/H}) \lesssim 8$, the predictions from the two models differ by $\sim1\,\mathrm{dex}$, while at higher metallicities they predict similar DtG ratios (c.f., the dashed and solid green lines in Figure \ref{fig:DustBuildupSimulations}). Observationally, however, this is an expensive task, given that low-mass, metal-poor galaxies in the EoR typically appear faint in \cii{} (e.g., \citealt{vallini2015,glazer2024}). 

A few of the REBELS-IFU galaxies with $12 + \log(\mathrm{O/H}) \gtrsim 8.5$ show $\log(\mathrm{DtG}) \lesssim -3$, which is in better agreement with the models from \citet{triani2020}. Their models predict dust build-up at $z\approx7$ proceeds mostly through SNe, and yield lower DtG and DtM ratios than the \citet{palla2024} and \citet{popping2017} models. On the other hand, our measurements appear inconsistent with the models from \citet{vijayan2019}, which yield DtG and DtM ratios that are $\sim0.5-1\,\mathrm{dex}$ lower than what we infer for the REBELS-IFU sample. While \citet{vijayan2019} predict ISM dust growth to be dominant already at $z\approx7$, their grain growth timescale may still be overestimated. Indeed, \citet{vijayan2019} show in their Appendix that further reducing the characteristic grain growth timescale would better reproduce observed dust masses at high redshift ($z\gtrsim5$) compared to their fiducial model, while retaining their agreement with dust mass measurements at Cosmic Noon and lower redshifts.

When comparing to the \citet{dayal2022} models, we find that they do not produce large numbers of galaxies with metallicities similar to the REBELS-IFU sources at $z\approx7$. This was previously noted by \citet{ucci2023} regarding the related \textsc{Astraeus} simulations, who ascribe this to the relatively low star formation efficiency in the models, in combination with perfect coupling of dust and metals in outflows, and potentially underestimated metallicities of inflowing gas. Nevertheless, an extrapolation of the fiducial \citet{dayal2022} model would seem to be in reasonable agreement with the high-DtG subset of our sample. Their `maximal' model produces DtM ratios significantly larger than what is observed either in the REBELS-IFU or $z\gtrsim6$ literature samples. While \citet{dayal2022} originally introduced this model to reproduce the high dust masses observed in some of the REBELS galaxies with low stellar masses, they added a caveat that the underlying assumptions are rather unrealistic (e.g., no dust destruction). As a result, the mismatch with the `maximal' models by \citet{dayal2022} is not particularly surprising.

\subsubsection{Dust-to-Stellar mass ratio}

Not all of the models introduced above provide predictions for the DtS ratio as a function of metallicity. Therefore, we compare only to the models by \citet{popping2017}, \citet{dayal2022} and \citet{palla2024} in the bottom panel of Fig.\ \ref{fig:DustBuildupSimulations}. While at high metallicities the REBELS-IFU sample agrees with both the fiducial and `no GG' model from \citet{popping2017}, the upper limit on the dust mass from REBELS-15 -- the source with the lowest metallicity in our sample -- is in better agreement with the fiducial scenario. Albeit at slightly higher redshift ($z\sim8.5$), the upper limit on the dust mass of S04950 is also consistent with their fiducial model, while being in tension with the model without ISM dust growth. The deep upper limits on the DtS ratios for GHZ2 at $z=12.33$ and GS-z14-0 at $z=14.18$ also suggest DtS ratios to be lower for metal-poor galaxies, although the \citet{popping2017} predictions are limited to $z < 9$, preventing a detailed comparison at such early epochs. 

Interestingly, a recent study by \citet{narayanan2024} suggests an evolutionary link between the blue and apparently dust-poor galaxies observed at $z\gtrsim10$ and the massive, dust-rich REBELS-IFU-like galaxies observed at later cosmic times. While \citet{narayanan2024} do not provide a relation between the DtS ratio and metallicity in their work, their models are successful at reproducing the population of galaxies with $-3 \lesssim \log(\mathrm{DtS}) \lesssim -2.5$ by $z\sim6-7$. They predict that the progenitors of these galaxies have a significantly lower $\log(\mathrm{DtS}) \sim -4$ at $z\gtrsim10$, while between $z\sim10$ to $z\sim7$ efficient ISM dust growth is responsible for rapidly building up their dust mass. This is in qualitative agreement with the high $\mathrm{DtS}$ ratios observed for the REBELS-IFU galaxies, compared to the stringent upper limits on the $\mathrm{DtS}$ ratios for the $z\gtrsim10$ galaxy population, although more data especially for these very high-redshift sources is essential to confirm this prediction. 

Returning to the models shown in Figure \ref{fig:DustBuildupSimulations}, we find that the REBELS-IFU data are also in good agreement with the dust-rich model from \citet{palla2024}, while their intermediate-dust model predicts DtS ratios that are $\sim0.5-1\,\mathrm{dex}$ lower than observed. However, neither model predicts strong variations in DtS with metallicity, which may be in slight tension with the aforementioned metal-poor galaxies in the literature at $z\approx8.5-14.2$. Finally, the REBELS-IFU sample is more metal-rich than the bulk of the galaxies in the \citet{dayal2022} models, and appear to agree better with (an extrapolation of) their `maximal model' in terms of the predicted DtS ratios. However, as mentioned, this model predicts far higher DtM ratios than inferred for the REBELS-IFU sample, and is thus overall inconsistent with our measurements. \\ 

In summary, the REBELS-IFU data are in best agreement with models predicting swift dust build-up by $z\approx7$ through a combination of supernova dust and rapid ISM dust growth (i.e., the fiducial and dust-rich models by \citealt{popping2017} and \citealt{palla2024}, respectively), or efficient supernovae alone (the `no grain growth' model by \citealt{popping2017}). Upon also including low-metallicity galaxies from the literature, we find that the fiducial model by \citet{popping2017} can best reproduce the observations, although we caution that this metal-poor comparison sample lies at higher redshifts than the REBELS-IFU sources. Nevertheless, reproducing the full suite of high-redshift observations appears to require a combination of modest SN yields to reproduce the upper limits on the DtS ratios of metal-poor galaxies, and rapid ISM dust growth to match the high DtG, DtM and DtS ratios of the REBELS-IFU sources.

Our conclusions do not change significantly if we adopt $O32$-based metallicities for our entire sample, as discussed in Appendix \ref{app:metallicity}. In fact, since this diagnostic yields slightly lower metallicities on average while leaving the DtG and DtS ratios unchanged, it further reinforces our conclusions that the dust contents of the REBELS-IFU sample are best reproduced by models predicting rapid dust build-up by $z\approx7$.

\section{Conclusions}\label{sec:conclusions}
We performed a detailed study of the dust, gas and metal contents of twelve galaxies at $6.5 \lesssim z \lesssim 7.7$ observed as part of the REBELS-IFU survey. Our sample benefits from \cii{} and dust continuum observations from the ALMA REBELS program \citep{bouwens2022,inami2022}, as well as metallicity measurements from \textit{JWST}/NIRSpec IFU observations (\citealt{rowland2025}; Stefanon et al.\ in preparation). We adopt dust masses for our targets obtained from multi-band ALMA observations (2 sources; \citealt{algera2024,algera2024b}), or from a single-band continuum datapoint assuming a fixed dust temperature of $T_\mathrm{dust} = 45\pm15\,\mathrm{K}$ (10 sources, 2 of which are upper limits). Moreover, we infer molecular gas masses from the \cii{} luminosity following the calibration by \citet{zanella2018}. Our key results are as follows:

\begin{itemize}
    \item We find that the REBELS-IFU sources have high gas fractions, ranging from $f_\mathrm{gas} \approx 0.73 - 0.96$. A subset of the sample shows gas fractions $f_\mathrm{gas} \sim 0.9$ while simultaneously possessing a high metallicity $12 + \log(\mathrm{O/H}) \gtrsim 8.0$. This is in tension with closed box evolutionary models, which should yield the maximum possible $f_\mathrm{gas}$ for a given metallicity. The most likely interpretation for this is that their gas masses are overestimated; if indeed the case, our dust-to-gas and dust-to-metal ratios will be larger than the fiducial values quoted in the following, which further strengthens our conclusion that dust build-up is highly efficient in massive galaxies at $z\approx7$ (see below).
    
    \item We measure a typical dust-to-gas ratio of $\log(\mathrm{DtG}) = -3.02 \pm 0.23$ for the dust-detected REBELS-IFU sample across a metallicity range of $7.9 \lesssim 12 + \log(\mathrm{O/H}) \lesssim 8.7$. This value does not show a strong dependence on metallicity, in contrast to observations at $z\approx0$ finding an approximately linear relation in this regime. This could plausibly be due to systematic uncertainties in our dust masses: considering only the five $z\approx6-7.3$ sources with accurate multi-band dust mass measurements (two from REBELS + three literature galaxies), we see evidence for an increasing trend between the DtG ratio and metallicity. If confirmed through multi-band ALMA observations of additional high-redshift galaxies, this would suggest the local scaling relation between dust-to-gas ratio and metallicity does not evolve significantly between $z\approx0-7$. 

    \item We measure dust-to-metal ratios spanning $-1.3 \lesssim \log(\mathrm{DtM}) \lesssim -0.2$ for the REBELS-IFU galaxies, which implies that $\sim5-36\,\%$ of all metals are locked up in dust grains in massive galaxies at $z\approx7$. 
    
    \item We determine dust-to-stellar mass ratios of $-2.7 \lesssim \log(\mathrm{DtS}) \lesssim -1.8$, in agreement with previous studies of $z\approx7$ galaxies. These values are consistent with those of dusty star-forming galaxies at cosmic noon, but are larger than those of local galaxies with similar metallicities as the REBELS-IFU sample. We do not find evidence for an increase in $\log(\mathrm{DtS})$ with metallicity, which could be taken as evidence for dust production through SNe alone, without the need for grain growth (though see the next bullet point). Assuming this to be the case, we infer an average supernova dust yield of $y_\mathrm{SN} = 0.52 \pm 0.28\,M_\odot\,\mathrm{SN}^{-1}$. 
 
    \item Our DtG, DtM and DtS ratios are in best agreement with theoretical models predicting rapid dust production in the early Universe. Using the REBELS-IFU sample alone, we cannot distinguish between the two key processes most likely responsible for building up this dust; SNe or efficient ISM dust growth. However, stringent upper limits on the dust masses of three metal-poor galaxies at higher redshifts ($z\gtrsim8.5$) may argue against very high SN yields, which suggests ISM dust growth to be important in building up the dust reservoirs of massive galaxies in the Epoch of Reionization.
    
\end{itemize}

Overall, our observations thus suggest that dust build-up is highly efficient in the early Universe. However, several open questions remain. While REBELS provides the first statistical insights into the dust contents of massive galaxies at $z\approx7$, our study is limited by sizable systematic uncertainties in both dust and gas masses. Multi-band ALMA observations are needed to better constrain the former, while high-resolution studies of \cii{} can be used to dynamically infer the total gas budget in light of uncertainties in the \cii{}-to-H$_2$ conversion factor. The REBELS sample is moreover relatively metal-enriched, and similar observations of lower-metallicity galaxies are essential to fully constrain the pathways of dust production and destruction at $z\approx7$. Finally, comparatively little is known about dust in the $z\gtrsim8$ Universe, and dedicated observing campaigns are needed to explore the gas and dust contents of galaxies at the earliest cosmic times.

\section*{Acknowledgements}
We thank the referee for their useful and constructive feedback that improved this work. This work was supported by NAOJ ALMA Scientific Research Grant Code 2021-19A (HSBA and HI). MS acknowledges support from the European Research Commission Consolidator Grant 101088789 (SFEER), from the CIDEGENT/2021/059 grant by Generalitat Valenciana, and from project PID2023-149420NB-I00 funded by MICIU/AEI/10.13039/501100011033 and by ERDF/EU. MA acknowledges support from ANID Basal Project FB210003 and and ANID MILENIO NCN2024\_112. RAAB acknowledges support from an STFC Ernest Rutherford Fellowship [grant number ST/T003596/1]. AF acknowledges support from the ERC Advanced Grant INTERSTELLAR H2020/740120. JH acknowledges support from the ERC Consolidator Grant 101088676 (``VOYAJ''). 

This paper makes use of the following ALMA data: \\ ADS/JAO.ALMA\#2017.1.01217.S, ADS/JAO.ALMA\#2019.1.01634.L, ADS/JAO.ALMA\#2021.1.00318.S, ADS/JAO.ALMA\#2021.1.01495.S, ADS/JAO.ALMA\#2022.1.01324.S, ADS/JAO.ALMA\#2022.1.01384.S. 

ALMA is a partnership of ESO (representing its member states), NSF (USA) and NINS (Japan), together with NRC (Canada), MOST and ASIAA (Taiwan), and KASI (Republic of Korea), in cooperation with the Republic of Chile. The Joint ALMA Observatory is operated by ESO, AUI/NRAO and NAOJ.

\section*{Data Availability}
The various quantities needed to reproduce the figures in this work are provided in Table \ref{tab:data}. Any further data underlying this article will be made available upon reasonable request to the corresponding author.



\bibliographystyle{mnras}
\bibliography{main} % if your bibtex file is called example.bib

%%%%%%%%%%%%%%%%%%%%%%%%%%%%%%%%%%%%%%%%%%%%%%%%%%

\appendix

\section{The $O32$ metallicity diagnostic}
\label{app:metallicity}

\begin{figure*}
    \centering
    \includegraphics[trim={0 0.5cm 0 0}, width=0.95\textwidth]{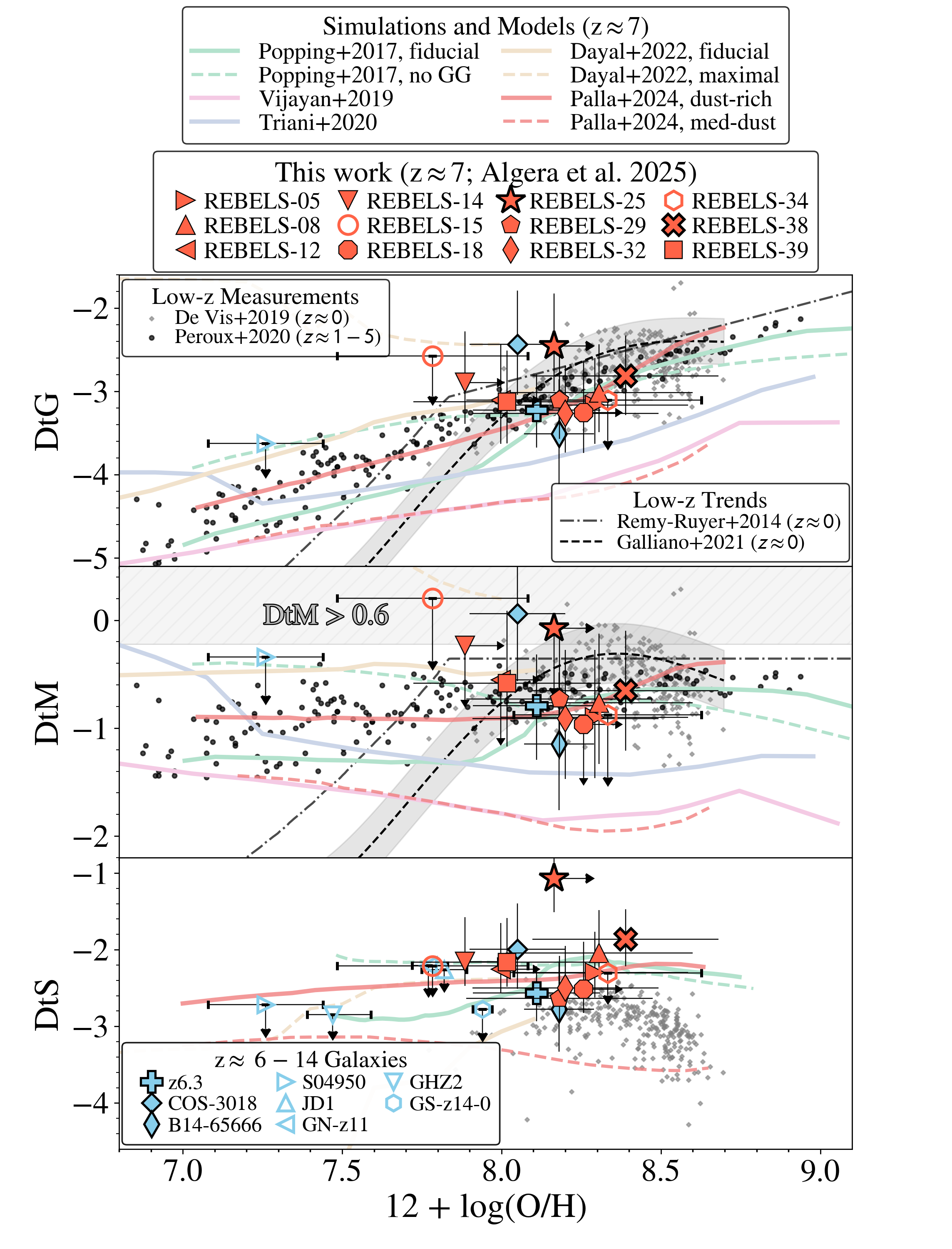}
    \caption{Same as Figures \ref{fig:DustBuildupObservations} and \ref{fig:DustBuildupSimulations} merged together, now consistently using $O32$ as a metallicity indicator for the REBELS-IFU sample. For the four galaxies at $z \gtrsim 7$, this yields lower limits on the oxygen abundance, given the absence of dust corrections. Overall, our conclusions do not change if we use $O32$-based metallicities in favor of our fiducial ones.}
    \label{fig:DustBuildUpAppendix}
\end{figure*}

As discussed in Section \ref{sec:methodsMetallicity}, we use different metallicity indicators for our sample, depending on the availability of dust corrections, and whether the relevant line ratio falls within a previously calibrated range. While this ensures we use the `most robust' metallicity diagnostic on a case-by-case basis, the various diagnostics each have their own systematic uncertainties (e.g., \citealt{kewley2008}). To address this problem, we merge Figures \ref{fig:DustBuildupObservations} and \ref{fig:DustBuildupSimulations} into one, and this time adopt a single, standardized metallicity indicator for the REBELS-IFU sample: $O32$ (Fig.\ \ref{fig:DustBuildUpAppendix}).

We note that for our four targets at $z\gtrsim7$, we cannot apply any dust corrections to the emission line fluxes as H$\alpha$ has redshifted out of the NIRSpec coverage. As a result, metallicities inferred from $O32$ are taken to be lower limits for these four galaxies, as \oii{}$_{3727,29}$ will be more attenuated than \oiii{}$_{5007}$. Consequently, this results in upper limits on the dust-to-metal ratios for these sources. 

As can be seen from Fig.\ \ref{fig:DustBuildUpAppendix}, our conclusions remain unchanged by adopting the $O32$ metallicity diagnostic: dust-to-gas and dust-to-stellar mass ratios are not affected, and in fact the lower average metallicities inferred using $O32$ yield a higher DtG at fixed metallicity compared to our fiducial analysis. As before, we find a more or less flat relation between DtG and metallicity, and therefore a tentative decrease between DtM and $12 + \log(\mathrm{O/H})$, albeit now across a more limited dynamic range.

\section{Literature samples with Dust, Gas and Metal Observations at $z>6$}
\label{app:literature}

While REBELS-IFU is the first homogeneously-selected sample with observations of cold gas (through \cii{}), dust (ALMA continuum) and metals (\textit{JWST}/NIRSpec) at $z \gtrsim 6$, a few individual galaxies at this epoch have amassed similarly detailed observations. We note that, even when such measurements all exist in the literature, in most cases they have not been analyzed in the context of scaling relations between dust and metals (the exceptions are presented in \citealt{heintz2023} and \citealt{valentino2024}), and we thus present this analysis for the first time here. 

We provide details on the literature sample below, in order of increasing redshift. For \cii{}-detected galaxies, we adopt $\alpha_\text{\cii{}} = 31\,M_\odot\,L_\odot^{-1}$ to maintain consistency with the REBELS-IFU sample. We note that no galaxies beyond $z > 8.5$ have currently been detected in \cii{} or dust emission (c.f., \citealt{tamura2019,bakx2020,fujimoto2024a}), and for galaxies at these early cosmic times we thus have no detailed information on their gas contents, and are furthermore limited to upper limits on their dust-to-stellar mass ratios. 

\textbf{RXCJ0600-z6.3} (hereafter z6.3; also known as the `Cosmic Grapes') is a gravitationally lensed galaxy at $z=6.07$ first identified through its bright \cii{} emission in the ALMA Lensing Cluster Survey \citep{fujimoto2021,laporte2021}. It was recently studied with \textit{JWST}, leading to an accurate measurement of its stellar mass [$\log(\mu M_\star / M_\odot) = 10.2 \pm 0.2$, where $\mu\approx 32$ is the magnification due to gravitational lensing] \citep{gimenez-arteaga2024} and metallicity [$12 + \log(\mathrm{O/H}) = 8.11 \pm 0.2$] \citep{fujimoto2024b}.\footnote{As we are interested in ratios of various quantities, such as the dust-to-stellar mass ratio, any uncertainties due to lensing magnification cancel out.} Moreover, \citet{valentino2024} used multi-band ALMA observations to robustly constrain the dust mass of z6.3 to be $\mu M_\mathrm{dust} = 4_{-2}^{+4} \times 10^7\,M_\odot$.

\textbf{COS-3018} is a Lyman-break galaxy at $z=6.85$, first identified through its high \oiii{}$_{4959,5007}$ + H$\beta$ equivalent width by \citet{smit2014}. It was first detected in \cii{} emission by \citet{smit2018}, and in ALMA Band 6 continuum emission by \citet{schouws2022}. Subsequently, \citet{witstok2022} determined its dual-band dust mass leveraging a stringent non-detection of its continuum emission in ALMA Band 8. For consistency with the REBELS-IFU sample analyzed in this work, we adopt the dust mass from \citet{algera2024} who refitted the SED of COS-3018 to find $\log(M_\mathrm{dust}/M_\odot) = 7.7_{-0.5}^{+0.6}$. Recently, \citet{scholtz2024} undertook a detailed study of COS-3018 using NIRSpec observations from the GA-NIFS program, finding the galaxy to consist of three separate components. \citet{scholtz2024} measure the total stellar mass of the system to be $\log(M_\star/M_\odot) = 9.7 \pm 0.1$, while its metallicity ranges from $12 + \log(\mathrm{O/H}) = 7.9 - 8.2$. Since we focus on the global properties of COS-3018 in this work, we adopt an average value of $12 + \log(\mathrm{O/H}) = 8.05 \pm 0.15$, with the uncertainty representing the observed spread between the components. 

\textbf{B14-65666}, also known as the `Big Three Dragons', is a UV-luminous merging galaxy system at $z=7.15$ first photometrically identified through ground-based imaging by \citet{bowler2014}, and subsequently studied in detail with both ALMA and later the \textit{JWST} in various works \citep{bowler2018,hashimoto2019,sugahara2021,sugahara2024,jones2024}. Relevant for our analysis is the \cii{} detection by \citet{hashimoto2019}, from which we obtain a gas mass estimate (see also \citealt{hashimoto2023}). B14-65666 is moreover continuum-detected in three ALMA bands \citep{sugahara2021}, which \citet{algera2024} re-fit to obtain a dust mass of $\log(M_\mathrm{dust}/M_\odot) = 7.0_{-0.5}^{+0.6}$. We adopt the stellar mass of B14-65666 from \citet{sugahara2024}, who measure $\log(M_\star/M_\odot) = 9.8 \pm 0.2$. Finally, \citet{jones2024} find that the global metallicity of B14-65666 ranges from $12 + \log(\mathrm{O/H}) = 8.07 - 8.28$, depending on which metallicity diagnostic is adopted. We simply adopt the average here with the uncertainty spanning the possible range, i.e., $12 + \log(\mathrm{O/H}) = 8.18 \pm 0.11$.

\textbf{S04950} is a gravitationally lensed galaxy at $z=8.50$ studied by \citet{heintz2023} and \citet{fujimoto2024a}. It was initially spectroscopically confirmed by \textit{JWST}/NIRSpec, which also yielded a metallicity measurement of $12 + \log(\mathrm{O/H}) = 7.26 \pm 0.18$ \citep{fujimoto2024a}. S04950 was thereafter targeted in both \cii{} and \oiii{}$_{88}$ emission with ALMA, yielding a tentative \cii{} detection and an upper limit on its dust mass of $\log(\mu M_\mathrm{dust} / M_\odot) < 6.0$ \citep{fujimoto2024a}. We moreover adopt the (magnified) stellar mass from their work of $\log(\mu M_\star/M_\odot) = 8.72_{-0.29}^{+0.31}$ (see also \citealt{gimenez-arteaga2023}).

\textbf{MACS1149-JD1}, hereafter JD1, is a gravitationally lensed galaxy at $z=9.11$ first spectroscopically confirmed through its \oiii{}$_{88}$ emission by \citet{hashimoto2018}. Deeper observations of the \oiii{} line were presented by \citet{tokuoka2022}, who also measured a sensitive upper limit on the dust continuum emission of JD1. They do not, however, provide an upper limit on the dust mass of JD1, and we therefore derive one assuming a temperature of $T_\mathrm{dust} = 45 \pm 15\,\mathrm{K}$ as we did for the REBELS-IFU sample to obtain $M_\mathrm{dust} < 1.6 \times 10^5\,M_\odot$ after accounting for gravitational magnification ($\mu = 10$, following \citealt{hashimoto2018}). Given that dust temperatures may increase with redshift (e.g., \citealt{sommovigo2022,mitsuhashi2024}), this is likely a conservative upper limit. \cii{} emission from JD1 was targeted but not detected by \citet{laporte2019}. We adopt the stellar mass measurement of JD1 by \citet{marconcini2024} of $\log(M_\star/M_\odot) = 7.47 \pm 0.05$, and use the metallicity measured by \citet{morishita2024} of $12 + \log(\mathrm{O/H}) = 7.82 \pm 0.07$. 

\textbf{GN-z11} is a UV-luminous Lyman-break galaxy at $z=10.60$ \citep{oesch2016,bunker2023}. Its \cii{} and underlying dust continuum emission were recently targeted by \citet{fudamoto2024} using the Northern Extended Millimeter Array (NOEMA), though neither was detected. As a result, \citet{fudamoto2024} inferred an upper limit of $\log(M_\mathrm{dust}/M_\odot) < 6.9$ on the dust mass of GN-z11. We adopt the stellar mass measured by \citet{tacchella2023} [$\log(M_\star/M_\odot) = 9.1_{-0.4}^{+0.3}$], while we use the metallicity from \citet{bunker2023} of $12 
+ \log(\mathrm{O/H}) = 7.77_{-0.05}^{+0.06}$. 

\textbf{GHZ2} is another UV-luminous galaxy at $z=12.33$, first spectroscopically confirmed by \textit{JWST} using both its NIRSpec and MIRI instruments \citep{castellano2024,zavala2024}. It was subsequently detected in \oiii{}$_{88}$ emission by \citet{zavala2024b} using ALMA, while the dust continuum in GHZ2 remained undetected. These sensitive dust continuum observations were recently presented by \citet{mitsuhashi2025}, though in our analysis we use the less sensitive ALMA continuum observations of GHZ2 from in \citet{bakx2023}. We furthermore adopt a stellar mass of $\log(M_\star/M_\odot) = 9.05_{-0.25}^{+0.10}$ from \citet{castellano2024} and a metallicity of $12 + \log(\mathrm{O/H}) = 7.47_{-0.08}^{+0.12}$ from \citet{calabro2024}. 

\textbf{JADES-GS-z14-0}, hereafter GS-z14-0, is one of the most distant known galaxies at $z=14.18$. Following its spectroscopic confirmation with NIRSpec \citep{carniani2024}, it was detected in \oiii{}$_{88}$ emission with ALMA \citep{carniani2024b,schouws2024}. These observations also yielded an upper limit on the dust mass of GS-z14-0 of $M_\mathrm{dust} < 10^6\,M_\odot$. We adopt the stellar mass [$\log(M_\star/M_\odot) = 8.78_{-0.10}^{+0.09}$] and metallicity [$12 + \log(\mathrm{O/H}) = 7.94 \pm 0.03$] of GS-z14-0 as measured by \citet{carniani2024b} (see also \citealt{helton2024}).

%%%%%%%%%%%%%%%%%%%%%%%%%%%%%%%%%%%%%%%%%%%%%%%%%%

% Don't change these lines
\bsp	% typesetting comment
\label{lastpage}
\end{document}